\documentclass[a4paper,usenames,dvipsnames,11pt]{article}
\pdfoutput=1

\usepackage{jheppub}
\usepackage{slashed}
\usepackage{mathrsfs,booktabs,multirow,tabularx}
\usepackage{stmaryrd}
\usepackage{xspace}
\usepackage{fancyvrb}
\usepackage[makeroom]{cancel}

\def\beq{\begin{equation}}
\def\eeq{\end{equation}}
\def\beqn{\begin{eqnarray}}
\def\eeqn{\end{eqnarray}}
\def\abs#1{\left|#1\right|}

\newcommand\sss{\scriptscriptstyle}

\newcommand\as{\alpha_{\sss S}}

\newcommand\aem{\alpha}
\newcommand{\gev}{\,\textrm{GeV}}

\newcommand\aNLO{{\sc\small MadGraph5\_aMC@NLO}}
\newcommand\aNLOs{{\sc\small MG5\_aMC}}
\newcommand\mf{{\sc\small MadFKS}}
\newcommand\ml{{\sc\small MadLoop}}
\newcommand\ct{{\sc\small CutTools}}
\newcommand\nin{{\sc\small Ninja}}
\newcommand\IREGI{{\sc\small IREGI}}

\newcommand\FJ{{\sc\small FastJet}}
\newcommand\lhapdfs{{\sc\small LHAPDF6}}

\newcommand\pt{p_{\sss T}}
\newcommand\kt{k_{\sss T}}
\newcommand\ptj{p_{\sss T}^{(j)}}
\newcommand\ptjo{p_{\sss T}^{(j_1)}}
\newcommand\ptjt{p_{\sss T}^{(j_2)}}
\newcommand\ptinc{p_{\sss T}^{\rm incl}}
\newcommand{\Ht}{H_{\sss T}}
\newcommand{\LOi}{{\rm LO}_i}
\newcommand{\LOo}{{\rm LO}_1}
\newcommand{\LOt}{{\rm LO}_2}
\newcommand{\LOth}{{\rm LO}_3}
\newcommand{\NLOi}{{\rm NLO}_i}
\newcommand{\NLOo}{{\rm NLO}_1}
\newcommand{\NLOt}{{\rm NLO}_2}
\newcommand{\NLOth}{{\rm NLO}_3}
\newcommand{\NLOf}{{\rm NLO}_4}

\newcommand{\bq}{\bar{q}}

\newcommand{\MSb}{\overline{\rm MS}}

\newcommand{\stepf}{\Theta}
\newcommand\muF{\mu_{\sss F}}
\newcommand\muR{\mu_{\sss R}}

\title{The complete NLO corrections to dijet hadroproduction}

\author[a]{R. Frederix,}
\author[b]{S. Frixione,}
\author[c]{V. Hirschi,}
\author[a,d]{D. Pagani,}
\author[e]{H.-S. Shao,}
\author[f,g]{M. Zaro}
\affiliation[a]{Physik Department T31, Technische Universit\"at M\"unchen, 
James-Franck-Str. 1,\\ D-85748 Garching, Germany}
\affiliation[b]{INFN, Sezione di Genova, Via Dodecaneso 33, I-16146, 
Genoa, Italy}
\affiliation[c]{SLAC, National Accelerator Laboratory
2575 Sand Hill Road, Menlo Park, CA 94025-7090, USA}
\affiliation[d]{Centre for Cosmology, Particle Physics and 
Phenomenology (CP3),\\Universit\'e Catholique de Louvain, 
B-1348 Louvain-la-Neuve, Belgium}
\affiliation[e]{TH Department, CERN, CH-1211 Geneva 23, Switzerland}
\affiliation[f]{Sorbonne Universit\'es, UPMC Univ. Paris 06, UMR 7589, LPTHE, 
 F-75005, Paris, France}
\affiliation[g]{CNRS, UMR 7589, LPTHE, F-75005, Paris, France}

\emailAdd{rikkert.frederix@tum.de}
\emailAdd{Stefano.Frixione@cern.ch}
\emailAdd{vahirsch@slac.stanford.edu}
\emailAdd{davide.pagani@tum.de}
\emailAdd{huasheng.shao@cern.ch}
\emailAdd{zaro@lpthe.jussieu.fr}

\abstract{We study the production of jets in hadronic collisions,
by computing all contributions proportional to $\as^n\aem^m$, with 
$n+m=2$ and $n+m=3$. These correspond to leading and next-to-leading 
order results, respectively, for single-inclusive and dijet observables
in a perturbative expansion that includes both QCD and electroweak effects.
We discuss issues relevant to the definition of hadronic jets in the
context of electroweak corrections, and present sample phenomenological
predictions for the 13-TeV LHC. We find that both the leading and 
next-to-leading order contributions largely respect the relative hierarchy 
established by the respective coupling-constant combinations.
}


\preprint{
\begin{flushright}
CERN-TH-2016-250\\
CP3-16-60\\
TUM-HEP-1074/16\\
SLAC-PUB-16890\\
\today
\end{flushright}
}

\begin{document}
\maketitle
\flushbottom

\section{Introduction\label{sec:intro}}
Jet production is a very common occurrence at high-energy hadron 
colliders; for example, at the 13-TeV LHC with an instantaneous
luminosity of ${\cal L}=10^{34}~{\rm cm}^{-2}{\rm s}^{-1}$, there are
several tens of thousands events per second that contain at least 
one jet with transverse momentum larger than 100~GeV. Such an abundance
allows experiments to carry out measurements affected by very small
statistical uncertainties, and thus to probe all corners of the phase
space in a multi-differential manner. At the same time, it constitutes
a severe problem for new-physics searches characterised by jet final states,
with the signal possibly swamped by Standard Model (SM) backgrounds.
This also applies to the easiest of cases, that of a dijet signature
(which is present in many beyond-the-SM scenarios, such as those that
feature heavy vector bosons, excited quarks, axigluons, Randall-Sundrum
gravitons, and so forth -- see e.g.~ref.~\cite{Harris:2011bh} for a 
review of experimental searches that focus on the dijet-mass 
spectrum), whose peak structure can be diluted by QCD 
effects or be difficult to study if at the border of the kinematically
accessible region. A well known example of the latter situation was the
high-$\pt$ excess reported by CDF~\cite{Abe:1996wy} in inclusive jet
events, that triggered a lot of interest owing to its being a possible
evidence of quark compositness, but that was ultimately entirely due to 
an SM effect. In particular, the PDFs used for computing the SM predictions 
to which the data had been compared were insufficiently constrained in
the $x$ region that dominated high-$\pt$ jet production, and the
uncertainties associated with their determination were unknown.

The case of the large transverse momentum excess at CDF typifies the
necessity of computing jet cross sections at the highest possible
accuracy in the SM. The largest of such cross sections is the dijet
one (which also gives the dominant contribution to single-inclusive
rates); we shall exclusively deal with it in this paper. 
Next-to-leading order (NLO) QCD results for inclusive and two-jet
distributions have been available since the early 1990's~\cite{Ellis:1990ek,
Aversa:1990uv,Ellis:1992en,Giele:1994gf}. The first complete next-to-NLO 
(NNLO) QCD predictions have appeared only very recently~\cite{Currie:2016bfm}.
As a rule of thumb based on the values of the respective coupling constants,
NNLO QCD effects (${\cal O}(\as^4)$) have the same numerical impact as the
so-called NLO ones in the electroweak (EW) theory (${\cal O}(\as^2\aem)$).
Partial pure-weak contributions to the latter had been computed in
refs.~\cite{Moretti:2006ea,Scharf:2009sp}, and the complete weak results
published in ref.~\cite{Dittmaier:2012kx}. The rationale for ignoring the 
NLO EW corrections of electromagnetic origin, which to the best of our 
knowledge have not been calculated so far, is the possible enhancement 
of weak contributions due to the growth of logarithmic terms of Sudakov
origin in certain regions of the phase space associated with large
scales~\cite{Ciafaloni:1998xg,Ciafaloni:2000df,Denner:2000jv,Denner:2001gw}, 
in particular at high transverse momenta. Incidentally, such 
Sudakov effects can also be responsible for large violations of the
natural hierarchy of QCD and EW corrections, with NLO EW ones 
becoming significantly larger than their NNLO QCD counterparts
and competitive with the NLO QCD results.

Motivated by the previous considerations, in this paper we present
the computation of {\em all} the leading and next-to-leading order
contributions to the dijet cross section in a mixed QCD-EW coupling 
scenario. In other words, we compute all the terms in the perturbative
series that factorise the coupling-constant combinations $\as^n\aem^m$, 
with $n+m=2$ (leading order, LO) and $n+m=3$ (NLO). Thus, we calculate
here for the first time the ${\cal O}(\as^2\aem)$ electromagnetic 
contribution, and the two NLO terms of ${\cal O}(\as\aem^2)$ and
${\cal O}(\aem^3)$. Our computations are carried out in the \aNLO\ 
framework~\cite{Alwall:2014hca} (\aNLOs\ henceforth), and are completely 
automated; this work therefore constitutes a further step in the validation
of the \aNLOs\ code, in a case that requires the subtraction of QED
infrared singularities which is significantly more involved than that
studied in ref.~\cite{Frixione:2015zaa}. We also take the opportunity
to discuss issues that arise when one defines jets in the presence of
final-state photon and leptons. 

This paper is organised as follows. In sect.~\ref{sec:setup} we outline
the contents of our computation and the general features of the framework in 
which it is performed. The problem of the definition of jets in the context 
of higher-order EW calculations is discussed in sect.~\ref{sec:jetdef}.
Phenomenological results for the LHC Run II are given in 
sect.~\ref{sec:results}. Finally, we present our conclusions 
in sect.~\ref{sec:conc}.

\section{Calculation setup\label{sec:setup}}
A generic observable in two-jet hadroproduction can be written
as follows:
\beqn
\Sigma_{jj}^{\rm (LO)}(\as,\aem)&=&
\as^2\,\Sigma_{2,0}
+\as\aem\,\Sigma_{2,1}
+\aem^2\,\Sigma_{2,2}
\nonumber\\*
&\equiv&
\Sigma_{\LOo}+\Sigma_{\LOt}+\Sigma_{\LOth}\,,
\label{SigB}
\\
\Sigma_{jj}^{\rm (NLO)}(\as,\aem)&=&
\as^3\,\Sigma_{3,0}
+\as^2\aem\,\Sigma_{3,1}
+\as\aem^2\,\Sigma_{3,2}
+\aem^3\,\Sigma_{3,3}
\nonumber\\*
&\equiv&
\Sigma_{\NLOo}+\Sigma_{\NLOt}+\Sigma_{\NLOth}+\Sigma_{\NLOf}\,,
\label{SigNLO}
\eeqn
at the LO and NLO respectively. The notation we adopt throughout
this paper is fully analogous to that of refs.~\cite{Alwall:2014hca,
Frixione:2014qaa,Frixione:2015zaa}. We refer the reader, in particular,
to ref.~\cite{Frixione:2014qaa} for a detailed discussion on the 
physical meaning of the terms that appear in eqs.~(\ref{SigB}) 
and~(\ref{SigNLO}), and the relevant terminology; we limit ourselves 
to recalling here that what is conventionally denoted by NLO QCD and NLO EW 
corrections can be identified with $\Sigma_{\NLOo}$ and $\Sigma_{\NLOt}$, 
respectively.

In our computation, $\Sigma_{jj}^{\rm (LO)}$ receives contributions
from all Feynman diagrams relevant to tree-level four-point Green
functions with external massless SM particles -- namely, light
quarks (including bottoms, since we work with five light flavours), gluons,
photons, and leptons\footnote{The reasons for this choice will be
discussed in sect.~\ref{sec:jetdef}.}. As far as $\Sigma_{jj}^{\rm (NLO)}$
is concerned, all one-loop four-point and tree-level five-point functions
with massless external legs contribute. Note that this implies that
while both real and virtual photons enter NLO corrections, $W^\pm$'s and
$Z$'s only appear as internal particles. Thus, what has been called HBR (for 
Heavy Boson Radiation) in refs.~\cite{Frixione:2014qaa,Frixione:2015zaa},
that is the contribution from tree-level diagrams that correspond to
the real emission of a $W^\pm$ or a $Z$ (and, in principle, one might
consider top-quark emissions, too) from a Born-level configuration, 
is not included in our results (incidentally, this is also the reason
why in the present case $\Sigma_{\rm NLO~EW}\equiv\Sigma_{\NLOt}$).
In fact, in order to consider HBR cross sections, one would need
either to possibly cluster a heavy vector boson together with other
massless particles when reconstructing jets (an option which is not
appealing from a physics viewpoint, given the procedure followed
by experiments), or to first decay any $W^\pm$ and $Z$ into a pair
of quarks or leptons. Having said that, we point out that \aNLOs\
can be used to simulate HBR contributions to dijet observables, and
that the corresponding calculations are fully independent of those
performed here.

All of the computations of the matrix elements mentioned above,
the renormalisation procedure, and the subtraction of the real-emission
infrared singularities (IR) are handled automatically by \aNLOs\ (with
a still-private version of the code). We remind the reader 
that \aNLOs\ makes use of the FKS method~\cite{Frixione:1995ms,
Frixione:1997np} (automated in the module \mf~\cite{Frederix:2009yq,
Frederix:2016rdc}) for dealing with IR singularities. The computations 
of one-loop amplitudes are carried out by switching dynamically between 
two integral-reduction techniques, OPP~\cite{Ossola:2006us} or 
Laurent-series expansion~\cite{Mastrolia:2012bu}, 
and TIR~\cite{Passarino:1978jh,Davydychev:1991va,Denner:2005nn}. 
These have been automated in the module \ml~\cite{Hirschi:2011pa}, which 
in turn exploits \ct~\cite{Ossola:2007ax}, \nin~\cite{Peraro:2014cba,
Hirschi:2016mdz}, or \IREGI~\cite{ShaoIREGI}, together with an in-house 
implementation of the {\sc OpenLoops} optimisation~\cite{Cascioli:2011va}.
Two remarks are in order here. Firstly, there is no element in the \aNLOs\ 
code that has been customised to compute dijet observables, in keeping with 
the general strategy that underpins the code. Secondly, although the papers 
cited above mostly treat explicitly the case of QCD corrections, \aNLOs\ has 
been constructed for being capable to handle other theories as well. For
what concerns the subtraction of real-emission singularities, the QED
case descends from the QCD one, with the most significant complications
in the context of automation due to bookkeeping (which understands the 
necessity of retaining independent control of the various $\Sigma_{k,q}$
terms). The underlying strategy has been outlined in 
sect.~2.4.1 of ref.~\cite{Alwall:2014hca}; the necessary
extensions to the code were chiefly carried out for the work of 
ref.~\cite{Frixione:2015zaa}, and further validated for the present
paper. As far as one-loop computations are concerned, \ml\ has been
completely overhauled in ref.~\cite{Alwall:2014hca} (see in particular
sects.~2.4.2 and~4.3 there), and it is since then that it is able to 
evaluate virtual amplitudes in theories other than QCD.

Finally, we point out that our simulations are entirely based on
a Monte Carlo integration of the short-distance subtracted cross
sections, that results in (weighted) events and their associated
counterevents. In particular, we do not use any factorised formulae
for the one-loop EW logarithmic corrections 
(see e.g.~ref.~\cite{Denner:2000jv}).

\section{Definition of jets\label{sec:jetdef}}
The prescription for the computation of a jet cross section, possibly 
in association with other objects, is unambiguous in perturbative QCD:
jets are composed of massless coloured particles (quarks and gluons), 
and this determines the nature of parton-level processes. Things become
more complicated as soon as one considers the first subleading higher-order
correction, i.e.~the electromagnetic one at the NLO. Among other things,
this entails the contribution of diagrams with an extra (w.r.t.~the
underlying Born configuration) real photon in the final state. In order
to have an IR-finite cross section, such a photon must be recombined
(at least in a suitable subset of the phase space) with nearby QCD 
partons to form a jet. However, this raises an issue when the jet is made
of a photon and a gluon: IR safety demands (consider the soft-gluon
limit) that there be an associated Born configuration in which a jet 
coincides with a photon. In other words, Born-level amplitudes must 
feature both QCD partons and photons (which in turn implies that one 
cannot limit oneself to considering only the leading, pure-QCD, 
Born contribution).

This does not really pose any problem: one must simply enlarge the set
of particles that can form jets at the level of short-distance cross 
sections (both at the leading and at higher orders), and include 
photons on top of light quarks and gluons; the resulting objects are
called {\em democratic jets}\footnote{Starting from the third-leading
NLO corrections (that scale like $(\aem/\as)^2$ w.r.t.~the leading,
pure-QCD, ones) it is necessary to include massless leptons as well
in the jet-clustering procedure. As far as we know, the term ``democratic''
applied to jets in a similar context has been used for the first time
in ref.~\cite{Glover:1993xc}.}.
The fact that a jet might be predominantly a non-hadronic quantity
is not surprising in a realistic experimental environment; for example,
in certain LHC analyses a jet is a spray of collimated particles with
up to 99\% of its energy of electromagnetic origin, of which up to
90\% can be carried by a single photon (see e.g.~refs.~\cite{Aad:2015nda,
Aad:2014rma,Aad:2014vwa,Aad:2013tea,Aad:2011fc,Khachatryan:2016mlc,
Khachatryan:2016wdh,Khachatryan:2016hkr,Khachatryan:2015luy,CMS:2014mna,
Chatrchyan:2014gia,Khachatryan:2014ika,Chatrchyan:2012bja} for a list of 
recent ATLAS and CMS papers approved as publications in the context of
jet physics). Having said that, fixed-order perturbation theory is
somehow pathological, precisely because a jet can coincide with
a photon. Although, as we shall show later, this situation is numerically
unimportant, it has motivated the introduction of procedures with 
the aim of getting rid of jets whose energy content is dominated by
a photon -- in this paper, we shall call such objects {\em photon jets}.
Recent examples can be found in refs.~\cite{Denner:2009gj,Denner:2014ina,
Kallweit:2014xda,Chiesa:2015mya}, that deal with NLO EW corrections to vector 
boson production in association with jets. The common feature of these 
procedures is the use of the photon energy (or of a related quantity, such 
as the transverse momentum), which is necessary to define the photon hardness,
and thus its relative contribution to that of the jet the photon belongs to.

Unfortunately, the photon energy is an ill-defined perturbative concept,
starting from the third-leading NLO correction (i.e.~$\Sigma_{\NLOth}$ 
in the case of dijet production). This can be easily seen by considering
a Born-level diagram with a final-state photon, and the real-emission
diagram obtained from the former by means of a $\gamma\to q\bq$ splitting:
by taking the $q\!\parallel\!\bq$ limit, one sees that the photon energy is
not an IR-safe quantity. 

In order to use photon degrees of freedom in an IR-safe way, the photon 
must be a physical final-state object (in other words, ``taggable''
or ``observable'').  For this to happen, the following rule must be obeyed:
\begin{itemize}
\item Photons can be considered as observable objects only if emerging 
from a fragmentation process. A photon that appears in a Feynman diagram 
has not been fragmented, and thus cannot be tagged.
\end{itemize}
A taggable photon is quite analogous to e.g.~a pion, which is described 
in perturbative QCD by means of a (non-perturbative) fragmentation process. 
As such, we shall have fragmentation functions that account for the 
long-distance process:
\beq
D^{(i)}_\gamma(z):\;\;\;\;i\longrightarrow\gamma\,,
\eeq
where $i$ is any massless particle that can fragment into a photon, and $z$ 
the fraction of the longitudinal momentum of $i$ carried by the photon.
Thus, the particle $i$ may be itself a photon, which is the most 
significant difference between the photon and the pion cases (since no pion 
can appear at the short-distance level). In particular, owing to the 
elementary nature of the photon, one will necessarily have~\cite{SCSF}:
\beq
D^{(\gamma)}_\gamma(z)=\left(A+B\aem+\cdots\right)\delta(1-z)+
\Delta D^{(\gamma)}_\gamma(z)\,,
\label{gammaFF}
\eeq
with $\Delta D^{(\gamma)}_\gamma(z)$ a regular function at $z\to 1$.
We point out that the ${\cal O}(\aem^0)$ $\delta(1-z)$ term in
eq.~(\ref{gammaFF}) is all one needs in the context of QCD computations
that feature final-state photons\footnote{Such a term corresponds to what
is usually called the direct contribution in pQCD calculations.}:
in that case, the difference between taggable photons 
and short-distance photons is irrelevant (and indeed
it is not necessary to introduce it). We also remark that it is perfectly
acceptable to have a process with both taggable and short-distance photons
in the final state; the degrees of freedom of the latter must be integrated 
over (as e.g.~in a jet-finding algorithm), while this is not necessary
(but still possible) for the former ones.

The scheme outlined above allows one to define a photon jet regardless
of the perturbative order in $\as$ and $\aem$ one is working at: for
example, a photon jet is any jet that contains a taggable photon with
energy $E_\gamma$ such that \mbox{$E_\gamma\ge z_{cut}E_j$}, with 
$E_j$ the jet energy and $z_{cut}$ a pre-defined constant.
However, in the context of a jet analysis what one is really interested
in is a ``hadronic'' jet, i.e.~a jet in which the content of EM energy
is smaller, not larger, than a given threshold (we shall call these
jets {\em anti-tagged jets} in this paper). This poses two problems.
Firstly, a photon can be anti-tagged not only if $E_\gamma<z_{cut}E_j$, 
but also if it simply escapes detection (which, for a fixed-order
theoretical calculation, is the case where the jet is made of quarks 
and gluons only, i.e.~one in which there is no photon). Secondly,
the anti-tagging condition creates a practical problem, because 
fragmentation functions can only be measured (if at all) for sufficiently 
large $z$'s. 

A possible solution to these problems employs again the idea of
photon jet. The starting point is the following identity (which is
the hadron-parton-duality unitary condition):
\beq
i\;=\;\sum_{h}D^{(i)}_h(z)+\ldots=
\sum_{h\ne\gamma}D^{(i)}_h(z)+D^{(i)}_\gamma(z)+\ldots\,.
\label{partonID}
\eeq
where the dots on the r.h.s.~generically denote power-suppressed terms.
In words: parton $i$ fragments into any ``hadrons'', which will be 
eventually clustered into a jet (note that parton $i$ can be dressed by the 
perturbative radiation of other massless particles -- these are understood
in the notation of eq.~(\ref{partonID})). In the rightmost side of 
eq.~(\ref{partonID}), the sum over parton-to-hadron fragmentation functions 
is split into the sum of a term that features all hadrons different from 
the photon, and of a parton-to-photon term. By neglecting the power-suppressed
terms we re-write eq.~(\ref{partonID}) as follows:
\beq
i=\sum_{h\ne\gamma}D^{(i)}_h(z)+D^{(i)}_\gamma(z)\stepf(z_{cut}-z)
+D^{(i)}_\gamma(z)\stepf(z-z_{cut})\,,
\label{partonID2}
\eeq
i.e.~we introduce tagging and anti-tagging conditions, which we can do
because the photon emerges from a fragmentation process, and thus is
taggable. Thence:
\beq
\sum_{h\ne\gamma}D^{(i)}_h(z)+D^{(i)}_\gamma(z)\stepf(z_{cut}-z)=
i-D^{(i)}_\gamma(z)\stepf(z-z_{cut})\,.
\label{partonID3}
\eeq
The l.h.s.~of eq.~(\ref{partonID3}) is what we want: the anti-tag jet
contribution. Unfortunately, neither of the terms that appear there can 
be reliably computed (for all $z$'s). Conversely, the r.h.s.~of that equation 
is just fine: the two terms there correspond to the fully-democratic jet
cross section and to the photon-tagged one. If eq.~(\ref{partonID3})
is iterated over all possible final-state partons, one ends up by
{\em defining} in a natural manner the anti-tag jet cross section
as the democratic cross section, minus all tagged-photon cross sections,
with the number of photons ranging from one to the maximum number of
jets compatible with the perturbative order considered. In formulae,
this can be expressed as follows:
\beq
d\sigma_{X;nj}^{({\rm antitag})}=d\sigma_{X;nj}^{({\rm dem})}
-\sum_{k=1}^n d\sigma_{X+k\gamma;\,nj}\,.
\label{defantitag}
\eeq
with $X$ any set of objects that have to be found in the final state
on top of $n$ jets (importantly, taggable photons may appear in
such a set). The first term on the r.h.s.~of eq.~(\ref{defantitag})
is the democratic jet cross section; no taggable photons are present,
except those possibly in $X$. Each of the $n$ cross sections that appear
in the second term on the r.h.s.~of eq.~(\ref{defantitag}) is constructed 
by using the same short-distance processes as those that contribute to the 
first term, and by fragmenting $k$ final-state quarks, gluons, and photons 
in all possible ways; $n$ jets are finally reconstructed. 
All $n+1$ terms on the r.h.s.~of eq.~(\ref{defantitag}) are finite
and IR safe, and can be computed independently of each other in 
perturbation theory.

What has been done so far for photons can essentially be repeated
in the case of massless leptons. The main difference is that a fermion
line cannot be made to disappear by splitting, and this implies that
there is a way to tag a lepton that is not viable in the case of photons.
Still, IR safety requires that such a tagging is performed on an object
which is not the (short-distance) lepton itself, but its dressed version:
this is nothing but a jet, typically constructed with a small aperture,
that contains one lepton and whatever extra radiation surrounds it.
Alternatively, one can follow the same procedure as for photons,
namely introduce parton-to-lepton fragmentation functions.
Either way, one arrives at the idea of taggable leptons, which can
be employed to define lepton jets; the anti-tag jet cross section
in the l.h.s.~of eq.~(\ref{defantitag}) is then defined by inserting
on the r.h.s.~subtraction terms relevant to the lepton-jet cross 
sections\footnote{The use of the physical lepton masses leads to
alternative approaches (see e.g.~ref.~\cite{Dittmaier:2008md}). These
typically feature large-logarithmic terms, that expose their IR
sensitivity and necessitate a careful treatment; we believe that they
are best avoided in the context of jet analyses and lepton-jet rejection.}.

The procedure outlined so far puts QCD and QED on a rather similar
footing. In particular, this implies that as far as EW corrections are
concerned all computations can be conveniently performed in an $\MSb$-like 
scheme (such as the $G_\mu$ or $\aem(m_Z)$ ones). We point out that this 
procedure naturally leads to the prescription usually adopted in NLO EW 
computations (see e.g.~ref.~\cite{Denner:1991kt}) that associates a factor 
$\aem(0)$ to each external (short-distance) photon: such a factor results 
from the RG evolution of the photon-to-photon fragmentation function, whose
$\delta(1-z)$ term acquires an overall factor $\aem(0)/\aem(Q)$~\cite{SCSF}.

\subsection{Photon-jet cross sections}
We now return to eq.~(\ref{defantitag}) in order to define the
photon-jet cross sections that appear in the second term on the
r.h.s.~of that equation, for the case of dijet hadroproduction we
are interested in. As was discussed above, a construction 
valid for all the $\as^n\aem^m$ combinations necessarily entails the 
use of fragmentation functions, whose knowledge is presently far
from being satisfactory (bar perhaps for the quark-to-photon one).

Therefore, we have to adopt a pragmatic solution; this amounts to defining 
the photon-jet cross sections only for those ${\cal O}(\as^n\aem^m)$ terms
for which the introduction of a fragmentation function can be bypassed;
for the other terms, the photon-jet cross sections will be set
equal to zero, and thus our anti-tag dijet cross section will coincide 
with the democratic one\footnote{We always cluster leptons democratically,
which is fully justified by the fact that their contributions are very 
subleading, and numerically completely negligible.}. We do this in
the following way. The photon-jet cross sections are defined by
using the isolated-photon cross sections for one and two photons, 
constructed identically to what one usually does in perturbative QCD, 
and whose final states are suitably clustered into jets (as we shall
specify later). This implies that the relevant perturbative orders
are the following:
\beqn
&&1\gamma:\;\;{\cal O}(\as\aem+\as^2\aem)\equiv
\Sigma_{\LOt}+\Sigma_{\NLOt}\,,
\label{1phiso}
\\
&&2\gamma:\;\;{\cal O}(\aem^2+\as\aem^2)\equiv
\Sigma_{\LOth}+\Sigma_{\NLOth}\,,
\label{2phiso}
\eeqn
for the one- and two-isolated-photon cross sections respectively. This 
is implicitly equivalent to setting the photon-to-photon fragmentation 
function equal to $\delta(1-z)$, i.e.~to neglecting the contribution to
it due to higher-order QED effects. The cross sections that correspond 
to eqs.~(\ref{1phiso}) and~(\ref{2phiso}) could still depend on 
quark-to-photon and gluon-to-photon fragmentation functions; in order
to avoid this, we choose to work with the smooth isolation prescription
of ref.~\cite{Frixione:1998jh}, which sets their contributions identically
equal to zero.
More in details, we have implemented the following procedure:
\begin{enumerate}
\item
find jets democratically;
\item
find isolated photons; they are defined following
ref.~\cite{Frixione:1998jh} (using transverse momenta), with the same 
cone aperture as for jets, and with $n_\gamma=\epsilon_\gamma=1$;
\item
loop over those photons: if a photon belongs to a jet, and it carries
more than 90\% of the $\pt$ of that jet, then flag the jet as a 
candidate photon jet;
\item
candidate photon jets are considered as proper photon jets
if and only if:
\begin{itemize}
\item there is exactly one isolated photon, and one computes
either $\Sigma_{\LOt}$ or $\Sigma_{\NLOt}$;
\item there are exactly two isolated photons, and one computes
either $\Sigma_{\LOth}$ or $\Sigma_{\NLOth}$;
\end{itemize}
\item each photon jet gives an entry to the histograms relevant to
single-inclusive observables. For dijet correlations, there is an
histogram entry for each pair of jets, at least one of which is
a photon jet\footnote{We point out that dijet correlations can be constructed
by using a subset of all possible two-jet pairings, and we choose in
sect.~\ref{sec:results} to consider only observables defined by means 
of the two hardest jets.}.
\end{enumerate}
\enlargethispage*{50pt}
There are many possible variants to items 1--5 above, but we believe
that all those that are consistent with the general ideas outlined 
before will give very similar numerical results. The most important
thing to bear in mind is that, regardless of the specific choices
made for the isolation procedure, one is guaranteed to get rid of
those configurations where a photon jet coincides with a photon,
which is the semi-pathological situation, peculiar of fixed-order
calculations, that one typically would like to avoid.

We point out that, with the choices made here, each photon jet will
coincide with a democratic jet (while the opposite is obviously
not true). Therefore, item 5 implies a local
and exact cancellation of the photon-jet contributions, if all the
computations relevant to the cross sections on the r.h.s.~of 
eq.~(\ref{defantitag}) are performed simultaneously (i.e.~during the
same run), which is what we do. This not only improves the numerical
stability of the results, but also resembles very closely any possible
experimental procedure that would reject jets with too high a content
of EM energy.

\section{Results\label{sec:results}}
We now turn to presenting our predictions for a variety of single-inclusive
and dijet observables that result from $pp$ collisions at a center of mass
energy of 13~TeV (LHC Run II). We refer the reader to eqs.~(\ref{SigB}) 
and~(\ref{SigNLO}) for the definitions of the LO ($\Sigma_{\LOi}$, 
$i=1,2,3$) and NLO ($\Sigma_{\NLOi}$, $i=1,2,3,4$) contributions to the 
cross section, respectively; here, we shall show different
linear combinations of these quantities. Jets are defined by means of
the $\kt$ algorithm~\cite{Catani:1993hr} with $D=0.7$, and reconstructed 
with \FJ~\cite{Cacciari:2011ma}; as a default, we present results relevant
to democratic jets, but also explicitly assess the effect of removing
photon jets, as discussed in sect.~\ref{sec:jetdef}. For all of the 
observables considered here the contribution of forward jets is discarded,
by imposing the constraint:
\beq
\abs{y}<2.8\,.
\eeq
We work in the 
five-flavour scheme (5FS) where all quarks, including the $b$, are massless; 
electrons, muons, and taus, collectively called leptons, are massless as 
well, while the vector boson masses and widths have been set as follows:
\beqn
&&m_W=80.419~\gev\,,\;\;\;\;\;\;
m_Z=91.188~\gev\,,
\\
&&\Gamma_W=2.09291~\gev\,,\;\;\;\;\;\;
\Gamma_Z=2.50479~\gev\,.
\eeqn
The CKM matrix is taken to be diagonal, and the complex-mass 
scheme~\cite{Denner:1999gp,Denner:2005fg} is employed throughout.
The PDFs are those of the NNPDF2.3QED set~\cite{Ball:2013hta},
extracted from \lhapdfs~\cite{Buckley:2014ana} with number 244600;
these are associated with
\beq
\as(m_Z)=0.118\,.
\eeq
We work in the $G_\mu$ EW scheme, where:
\beq
G_\mu = 1.16639\cdot 10^{-5}\;\;\;\;\longrightarrow\;\;\;\;
\frac{1}{\aem}=132.507\,.
\label{aem}
\eeq
The central values of the renormalisation ($\muR$) and factorisation 
($\muF$) scales are both equal to:
\beq
\mu_0=\frac{\Ht}{2}\equiv 
\frac{1}{2}\sum_i \pt(i)\,,
\label{scref}
\eeq
where the sum runs over all final-state particles. The theoretical 
uncertainties due to the $\muR$ and $\muF$ dependencies have been 
evaluated by varying these scales independently in the range:
\beq
\frac{1}{2}\mu_0\le\muR,\muF\le 2\mu_0\,,
\label{scalevar}
\eeq
and by taking the envelope of the resulting predictions. The scale
dependence of $\aem$ is ignored, and the systematics associated with
the variations in eq.~(\ref{scalevar}) is evaluated by means of the
exact reweighting technique introduced in ref.~\cite{Frederix:2011ss}.
Reweighting is also employed for the computation of PDF uncertainties,
with individual weights combined according to the NNPDF 
methodology~\cite{Ball:2008by}. We report the 68\% CL symmetric interval 
(that is the one that contains only 68 replicas out of a total
of a hundred; this is done in order to avoid the problem of outliers, 
which is severe in this case owing to the photon PDF~\cite{Ball:2013hta}).
Finally, we note that the NNPDF2.3 set adopts a variable-flavour-number 
scheme. For scales larger than the top mass, this scheme is equivalent to 
the six-flavour one (6FS). Since the hard matrix elements are evaluated in 
the 5FS, the impact of the sixth flavour has to be removed from the running 
of $\as$ and from the DGLAP evolution of the PDFs. This corresponds to 
adding to the NLO 6FS-PDF cross section the following quantity:
\beq
\as\frac{T_F}{3\pi} \sum_{i,k} 
\left[
n_g^{(i,k)} \log\left( \frac{\muF^2}{m_t^2} \right) \stepf(\muF-m_t)- 
b^{(i,k)} \log\left( \frac{\muR^2}{m_t^2} \right) \stepf(\muR-m_t)
\right]\Sigma_{{\rm LO}_{ik}}\,.
\eeq
Here, $n_g^{(i,k)}$ and $b^{(i,k)}$ are the number of initial-state gluons 
and the power of $\as$ in $\Sigma_{{\rm LO}_{ik}}$, respectively, with $k$
numbering the individual partonic channels that contribute to 
$\Sigma_{\LOi}$. The interested reader can find more details in 
ref.~\cite{Cacciari:1998it} or in sect.~IV.2.2 of ref.~\cite{Badger:2016bpw}.

\begin{figure}[!ht]
\vskip -1.0truecm
\begin{center}
  \includegraphics[width=0.99\textwidth]{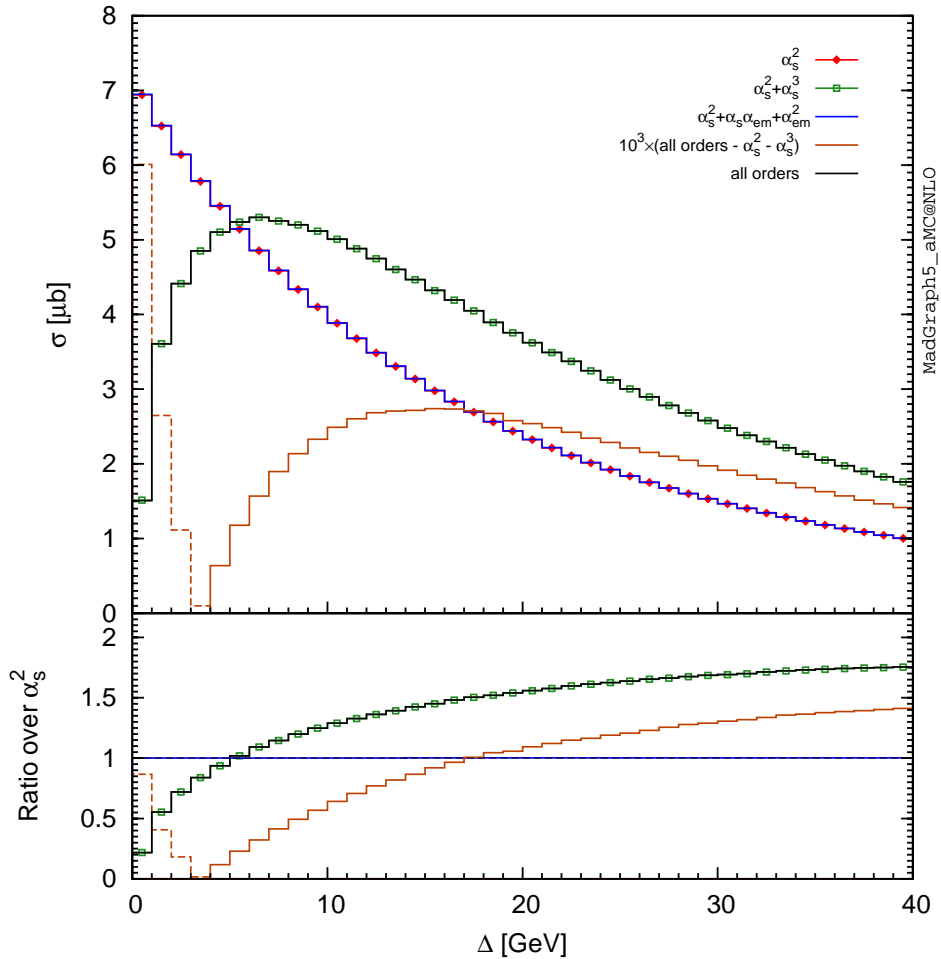}
\end{center}
\vskip -5.5truecm
\caption{\label{fig:Delta} 
Total dijet cross section as a function of $\Delta$, according to
the definition given in eq.~(\ref{Deltadef}).
}
\end{figure}

In order to determine which transverse-momentum cuts are sensible in an
NLO computation, we follow the procedure of ref.~\cite{Frixione:1997ks}
and present in fig.~\ref{fig:Delta} the total dijet cross section as 
a function of $\Delta$, according to the following definition:
\beq
\sigma(\Delta)=\sigma\left(\ptjo\ge 60~{\rm GeV}+\Delta,\;
\ptjt\ge 60~{\rm GeV}\right)\,,
\label{Deltadef}
\eeq
with $\ptjo$ and $\ptjt$ the transverse momentum of the hardest
and second-hardest jet, respectively. In other words, $\Delta$ 
measures the asymmetry between the $\pt$ cuts imposed on the two 
hardest jets, having assumed the transverse momentum of the second-hardest
jet to be larger than 60~GeV. Such a value is arbitrary, and is chosen
as typical of LHC jet analyses; we point out that its impact on the pattern
of the dijet cross section dependence upon $\Delta$ is negligible
(within a reasonable range).
There are five curves in the main frame of fig.~\ref{fig:Delta}.
The red histogram overlaid with full diamonds represents the $\Sigma_{\LOo}$ 
contribution, while the blue one corresponds to the sum of all of the 
LO contributions, $\Sigma_{jj}^{\rm (LO)}$. The green histogram overlaid 
with open boxes is the sum $\Sigma_{\LOo}+\Sigma_{\NLOo}$, i.e.~of the 
leading terms (pure QCD) at the LO and NLO; the black histogram is the sum 
of all of the LO and NLO contribution, 
\mbox{$\Sigma_{jj}^{\rm (LO)}+\Sigma_{jj}^{\rm (NLO)}$}, and is
denoted by ``all orders''.
Finally, the brown curve represents the sum of all LO and NLO contributions,
bar the pure QCD ones ($\Sigma_{\LOo}$ and $\Sigma_{\NLOo}$); in order for it
to fit into the frame of the figure, this histogram has been rescaled by a
factor of $10^3$.  In the region where the latter curve is displayed with a
dashed pattern, the cross section is negative, and thus what is represented is
its absolute value; this convention will be used throughout this section. The
lower panel in fig.~\ref{fig:Delta} presents the ratios of the results shown
in the main frame, over the $\Sigma_{\LOo}$ prediction.

As is explained in detail in ref.~\cite{Frixione:1997ks}, the dijet cross 
section behaves in a pathological manner for small $\Delta$ values at the NLO, 
owing to the presence of large $\log\Delta$ terms. Given the definition in
eq.~(\ref{Deltadef}), one would expect a monotonically increasing rate
for $\Delta\to 0$. This is indeed the behaviour of the LO results
(red-with-diamonds and blue histograms), while the NLO ones actually
decrease as $\Delta=0$ is approached. Figure~\ref{fig:Delta} therefore 
helps decide which value of $\Delta$ is appropriate in order to carry out
sensible NLO computations. Inspection of the plot suggests to set 
$\Delta\gtrsim 20$~GeV -- for such values, the three NLO predictions
are still monotonically growing. In order to be definite, we shall 
thus impose
\beq
\ptjo\ge 80~{\rm GeV}\,,\;\;\;\;\;\;\;\;\ptjt\ge 60~{\rm GeV}
\label{ptcuts}
\eeq
in our simulations for dijet correlations, while for single-inclusive
distributions we impose
\beq
\ptj\ge 60~{\rm GeV}\,.
\label{ptcutSI}
\eeq
There are a couple of further observations relevant
to fig.~\ref{fig:Delta}. Firstly, the full LO and NLO results (blue
and black histograms, respectively) are extremely close (but not identical,
although that is hard to see directly from the plot) to their leading, 
pure-QCD, counterparts (red-with-diamonds and green-with-boxes histograms, 
respectively). This is the well-known fact that EW contributions
are negligible as far as dijet rates are concerned, their effects being
manifest only in certain phase-space regions characterised by large scales
and that contribute little to total cross sections. Secondly, it appears
that the impact of $\log\Delta$ terms is larger when the pure-QCD
contributions are not included (the peak of the brown histogram occurs
at a much larger $\Delta$ value than that relevant to the two other
NLO results). This suggests that a conservative choice of $\Delta$
(similar to or even more stringent than that of eq.~(\ref{ptcuts})) 
is recommended where EW effects are particularly prominent.

\begin{figure}[!ht]
\vskip -1.0truecm
\begin{center}
  \includegraphics[width=0.99\textwidth]{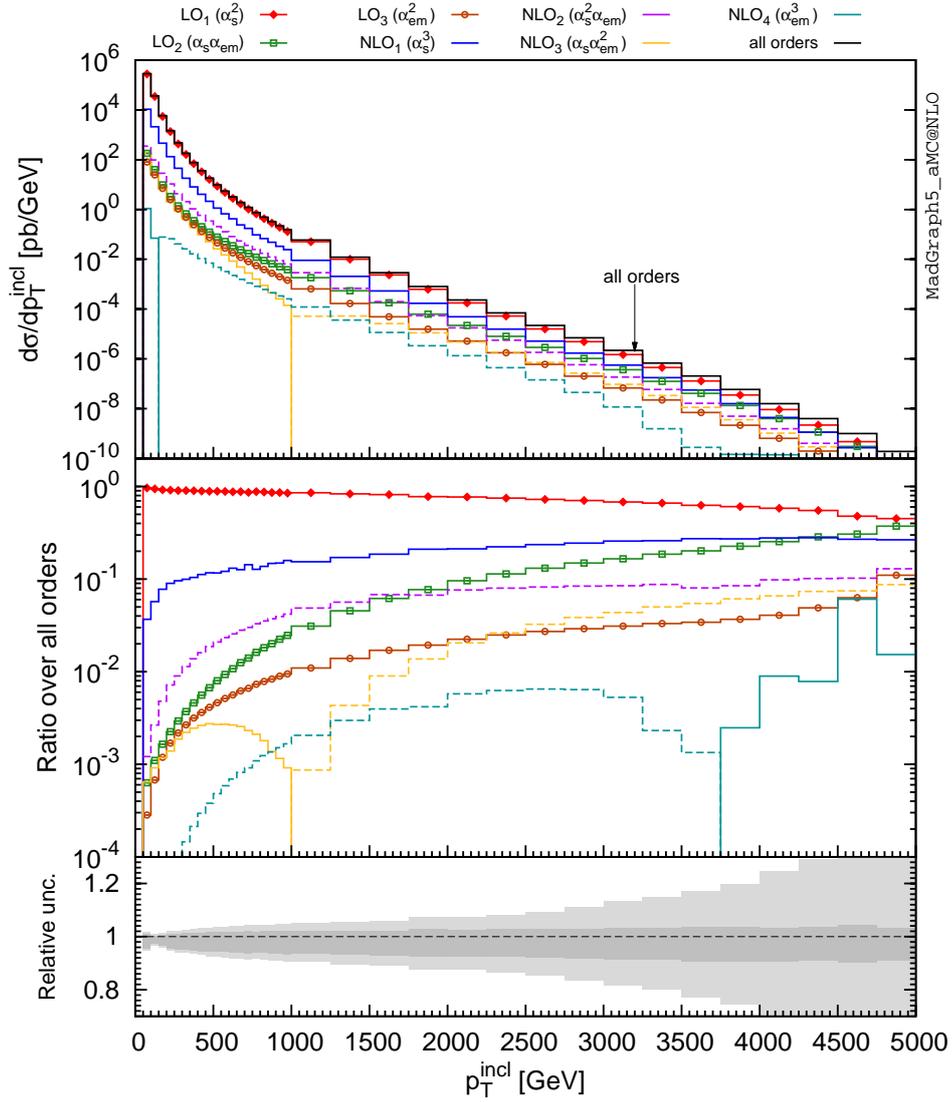}
\end{center}
\vskip -3.5truecm
\caption{\label{fig:pt1} 
Single-inclusive transverse momentum.
}
\end{figure}
\begin{figure}[!ht]
\vskip -1.0truecm
\begin{center}
  \includegraphics[width=0.99\textwidth]{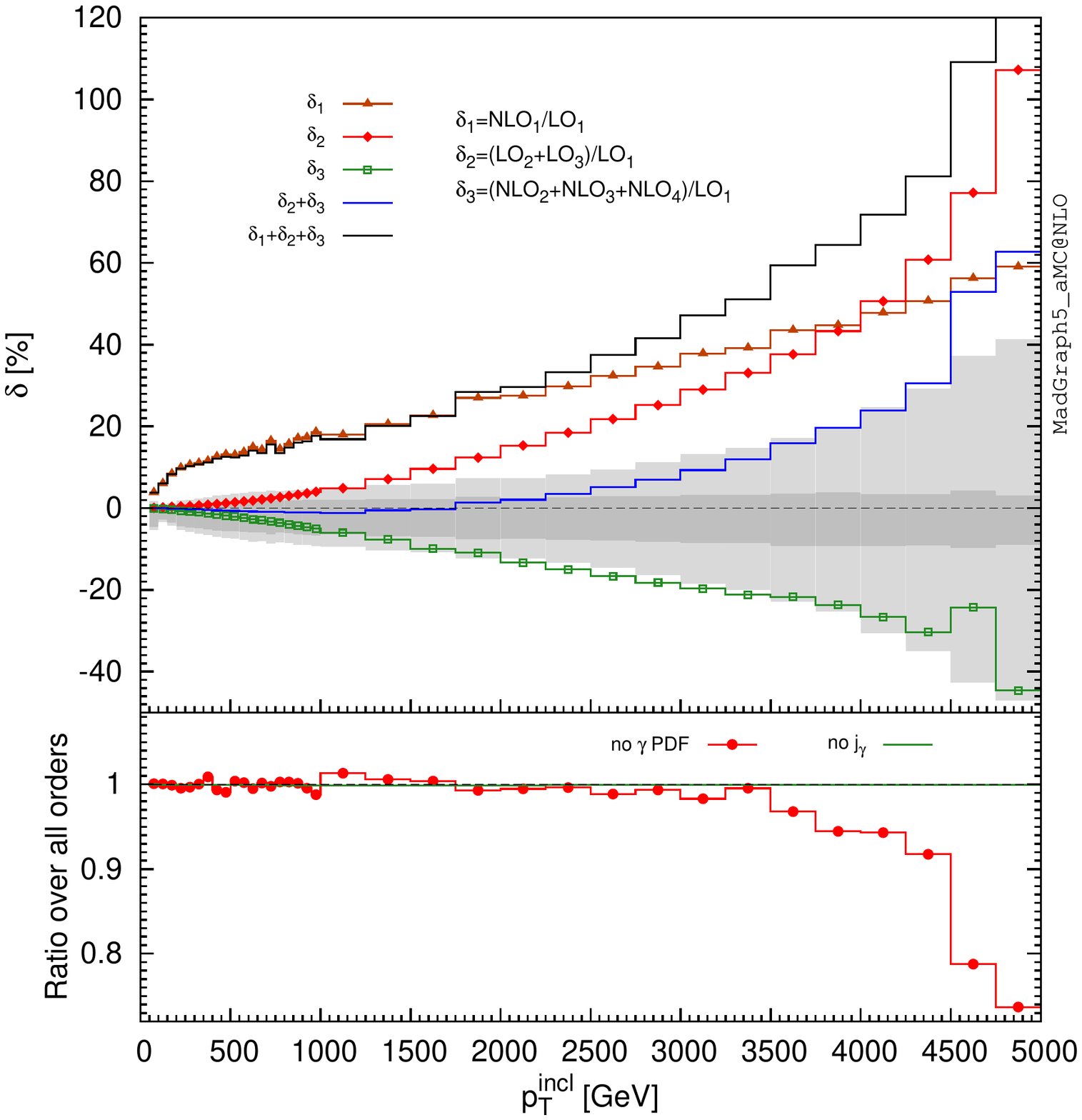}
\end{center}
\vskip -5.5truecm
\caption{\label{fig:pt2} 
Single-inclusive transverse momentum.
}
\end{figure}
\begin{figure}[!ht]
\vskip -1.0truecm
\begin{center}
  \includegraphics[width=0.99\textwidth]{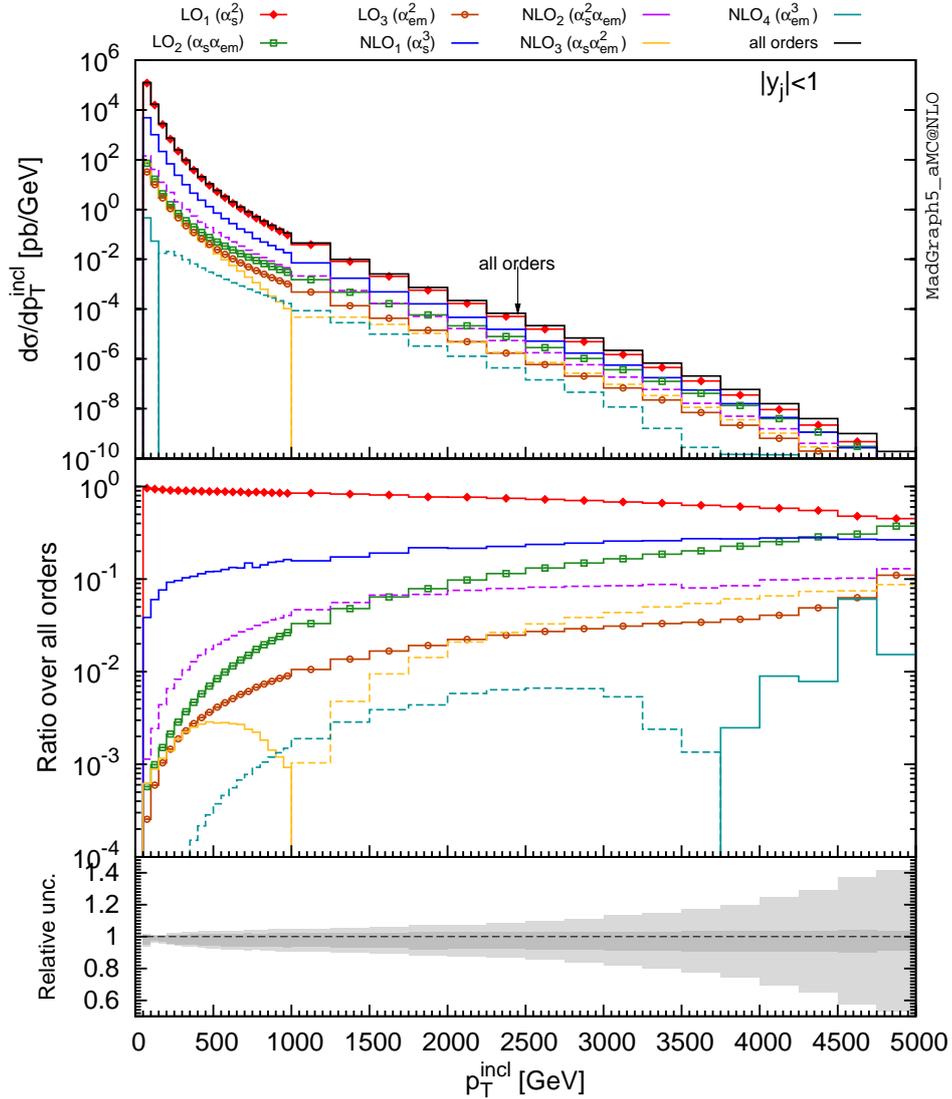}
\end{center}
\vskip -3.5truecm
\caption{\label{fig:pt1A} 
Single-inclusive transverse momentum, for $\abs{y}\le 1$.
}
\end{figure}
\begin{figure}[!ht]
\vskip -1.0truecm
\begin{center}
  \includegraphics[width=0.99\textwidth]{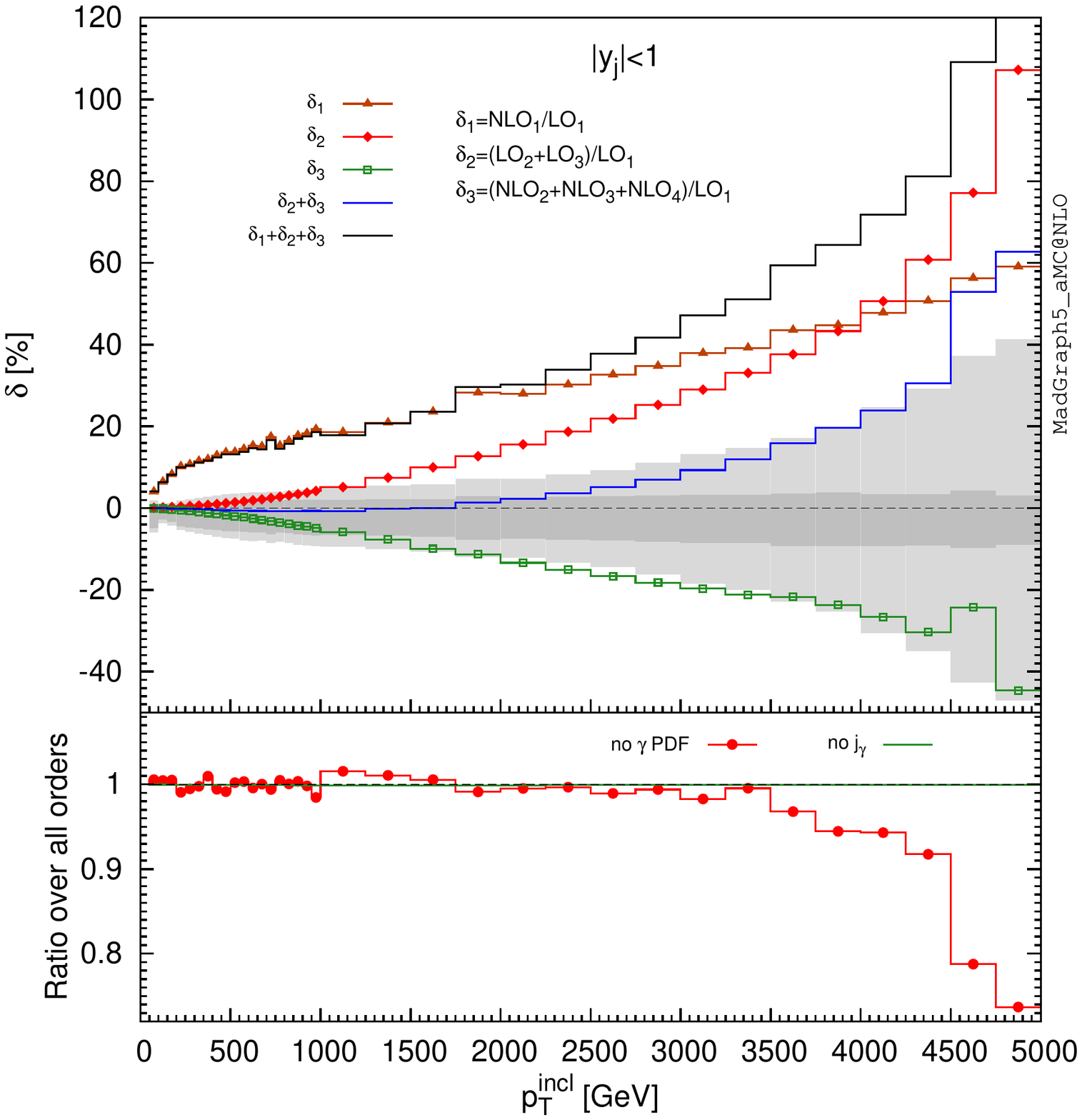}
\end{center}
\vskip -5.5truecm
\caption{\label{fig:pt2A} 
Single-inclusive transverse momentum, for $\abs{y}\le 1$.
}
\end{figure}
\begin{figure}[!ht]
\vskip -1.0truecm
\begin{center}
  \includegraphics[width=0.99\textwidth]{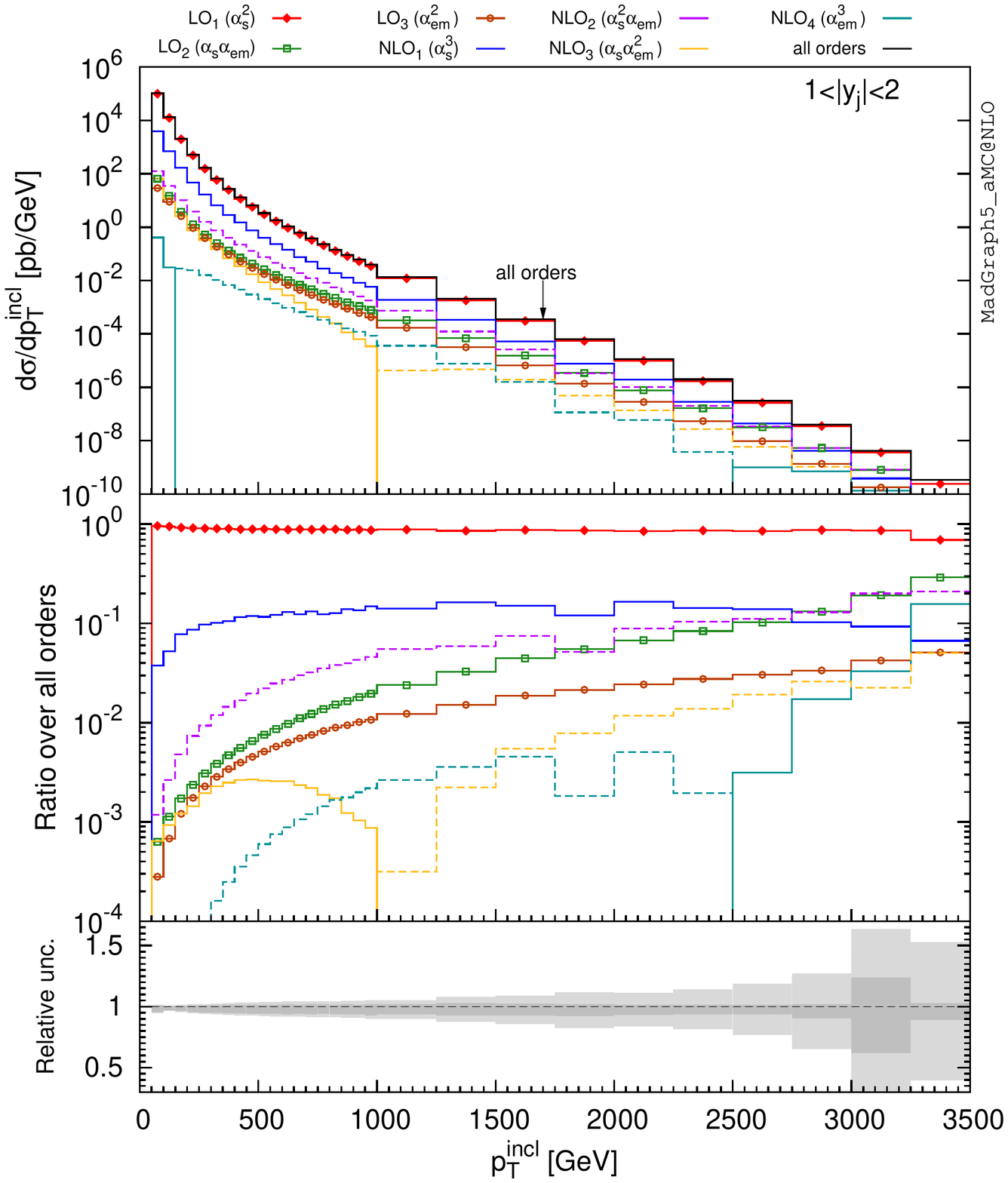}
\end{center}
\vskip -3.5truecm
\caption{\label{fig:pt1B} 
Single-inclusive transverse momentum, for $1<\abs{y}\le 2$.
}
\end{figure}
\begin{figure}[!ht]
\vskip -1.0truecm
\begin{center}
  \includegraphics[width=0.99\textwidth]{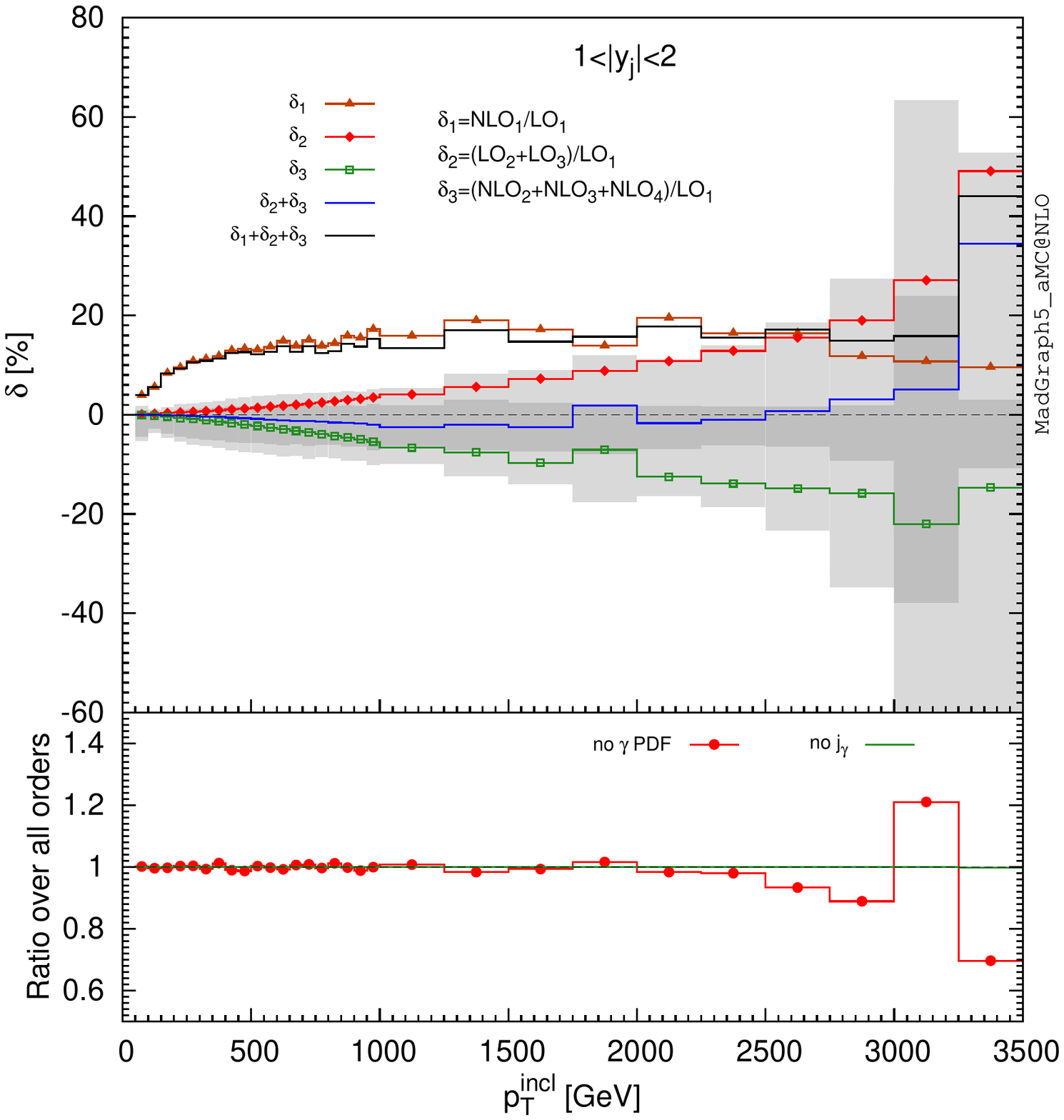}
\end{center}
\vskip -5.5truecm
\caption{\label{fig:pt2B} 
Single-inclusive transverse momentum, for $1<\abs{y}\le 2$.
}
\end{figure}
\begin{figure}[!ht]
\vskip -1.0truecm
\begin{center}
  \includegraphics[width=0.99\textwidth]{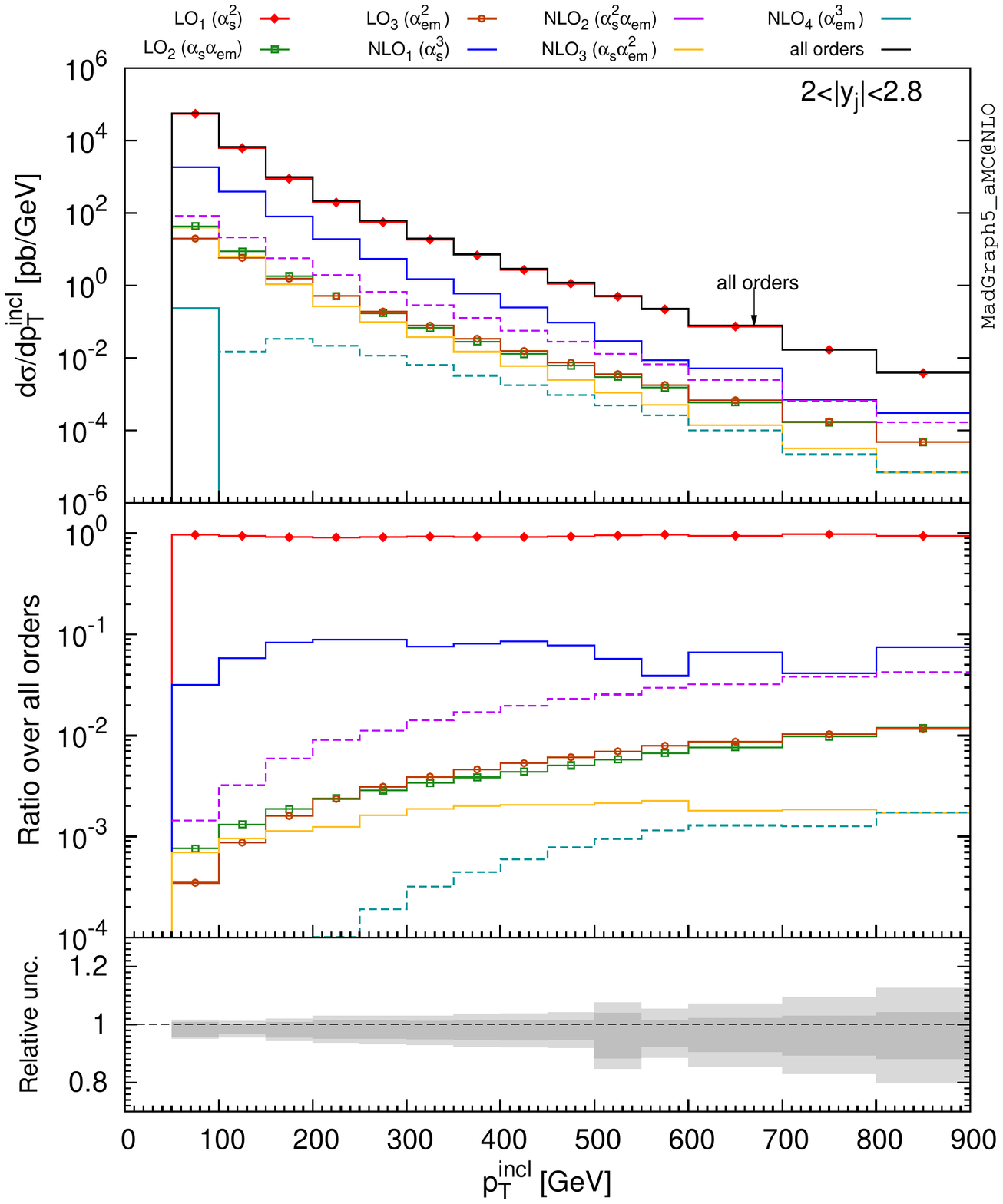}
\end{center}
\vskip -3.5truecm
\caption{\label{fig:pt1C} 
Single-inclusive transverse momentum, for $2<\abs{y}\le 2.8$.
}
\end{figure}
\begin{figure}[!ht]
\vskip -1.0truecm
\begin{center}
  \includegraphics[width=0.99\textwidth]{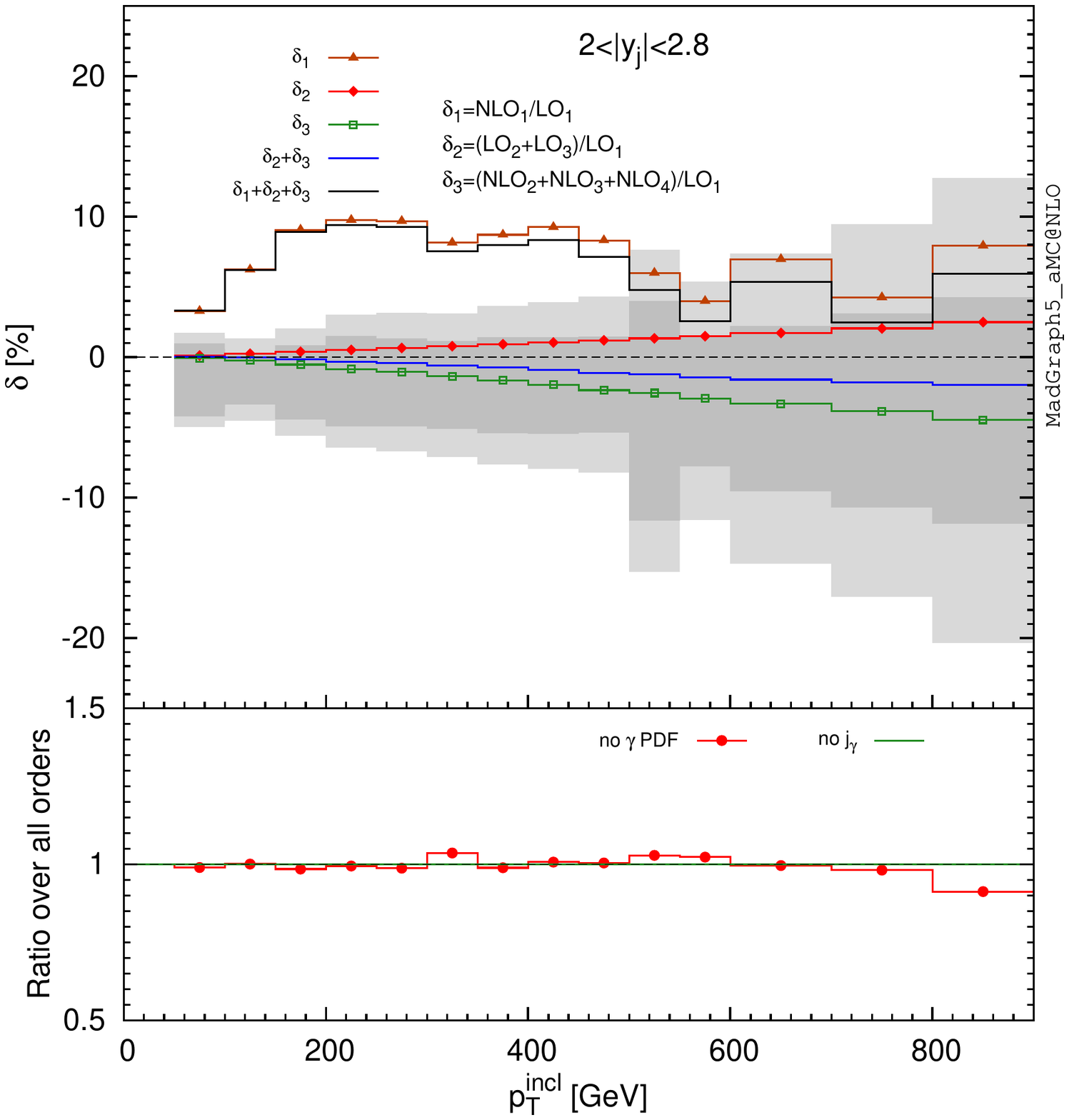}
\end{center}
\vskip -5.5truecm
\caption{\label{fig:pt2C} 
Single-inclusive transverse momentum, for $2<\abs{y}\le 2.8$.
}
\end{figure}
\begin{figure}[!ht]
\vskip -1.0truecm
\begin{center}
  \includegraphics[width=0.99\textwidth]{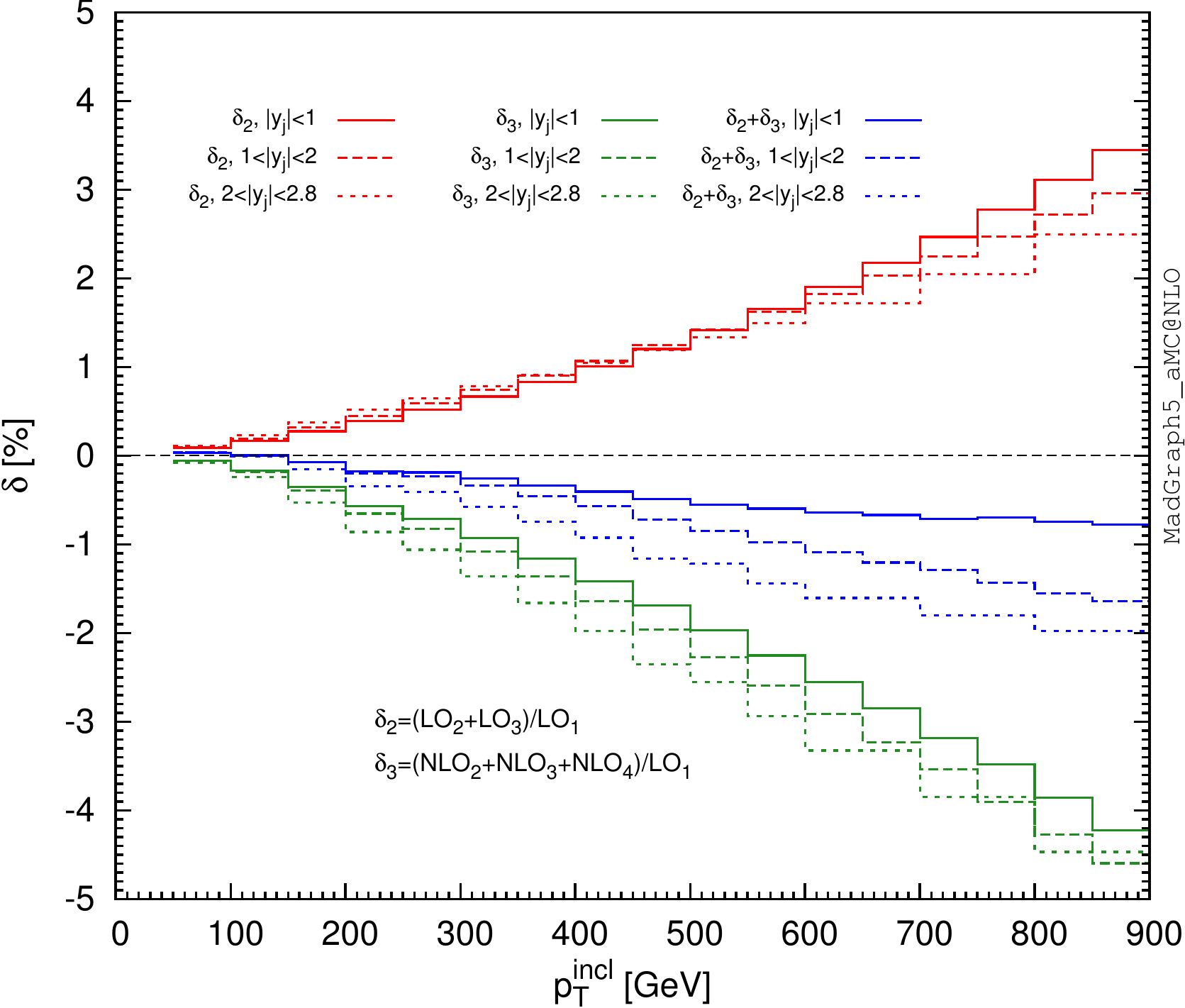}
\end{center}
\vskip -0.3truecm
\caption{\label{fig:pt2all} 
Single-inclusive transverse momentum; $\delta_2$ and $\delta_3$ predictions
for the three rapidity regions already considered in figs.~\ref{fig:pt2A},
\ref{fig:pt2B}, and~\ref{fig:pt2C}.
}
\end{figure}
\begin{figure}[!ht]
\vskip -1.0truecm
\begin{center}
  \includegraphics[width=0.99\textwidth]{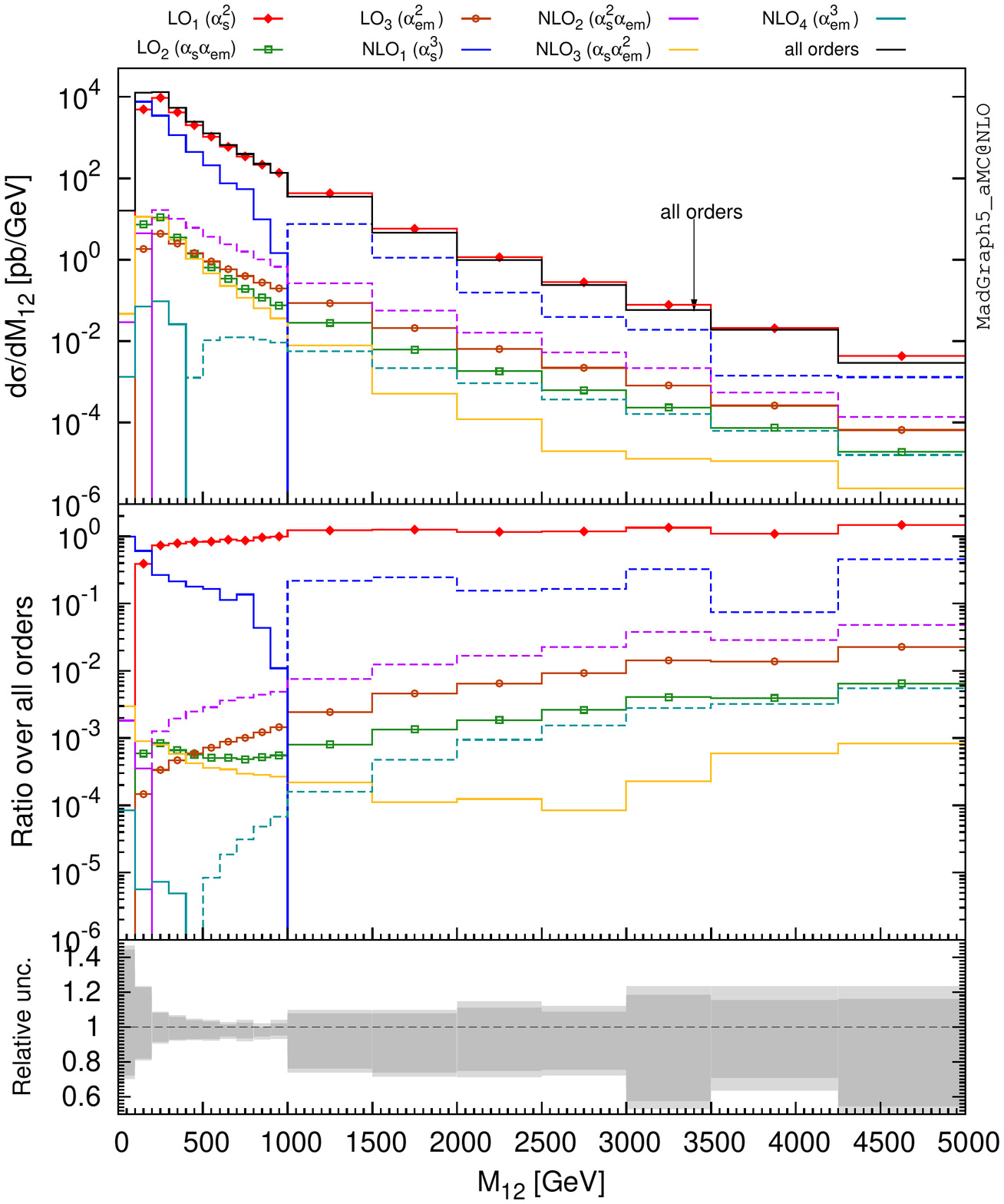}
\end{center}
\vskip -3.5truecm
\caption{\label{fig:M121} 
Invariant mass of the hardest jet pair.
}
\end{figure}
\begin{figure}[!ht]
\vskip -1.0truecm
\begin{center}
  \includegraphics[width=0.99\textwidth]{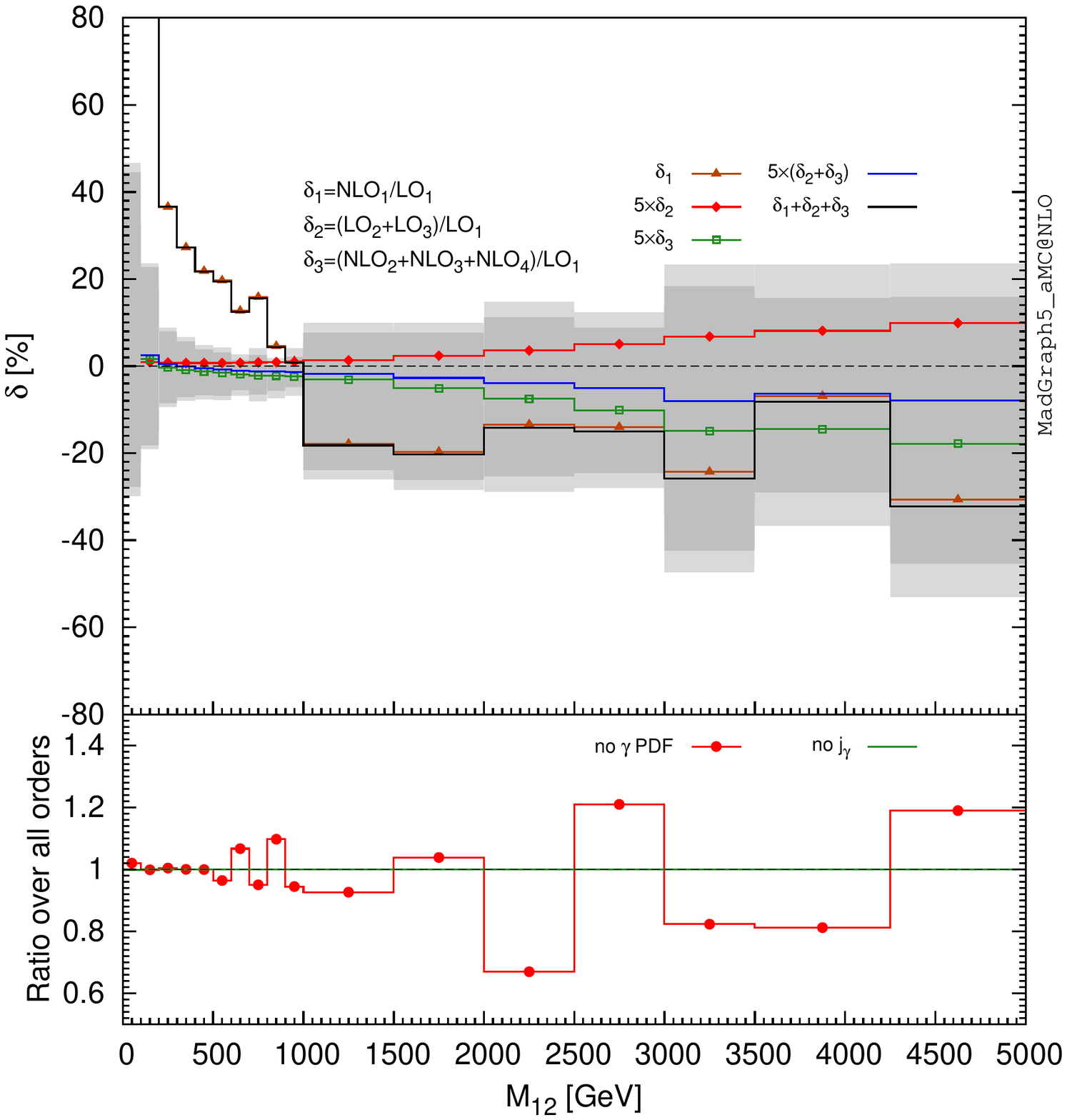}
\end{center}
\vskip -5.5truecm
\caption{\label{fig:M122} 
Invariant mass of the hardest jet pair.
}
\end{figure}
\begin{figure}[!ht]
\vskip -1.0truecm
\begin{center}
  \includegraphics[width=0.99\textwidth]{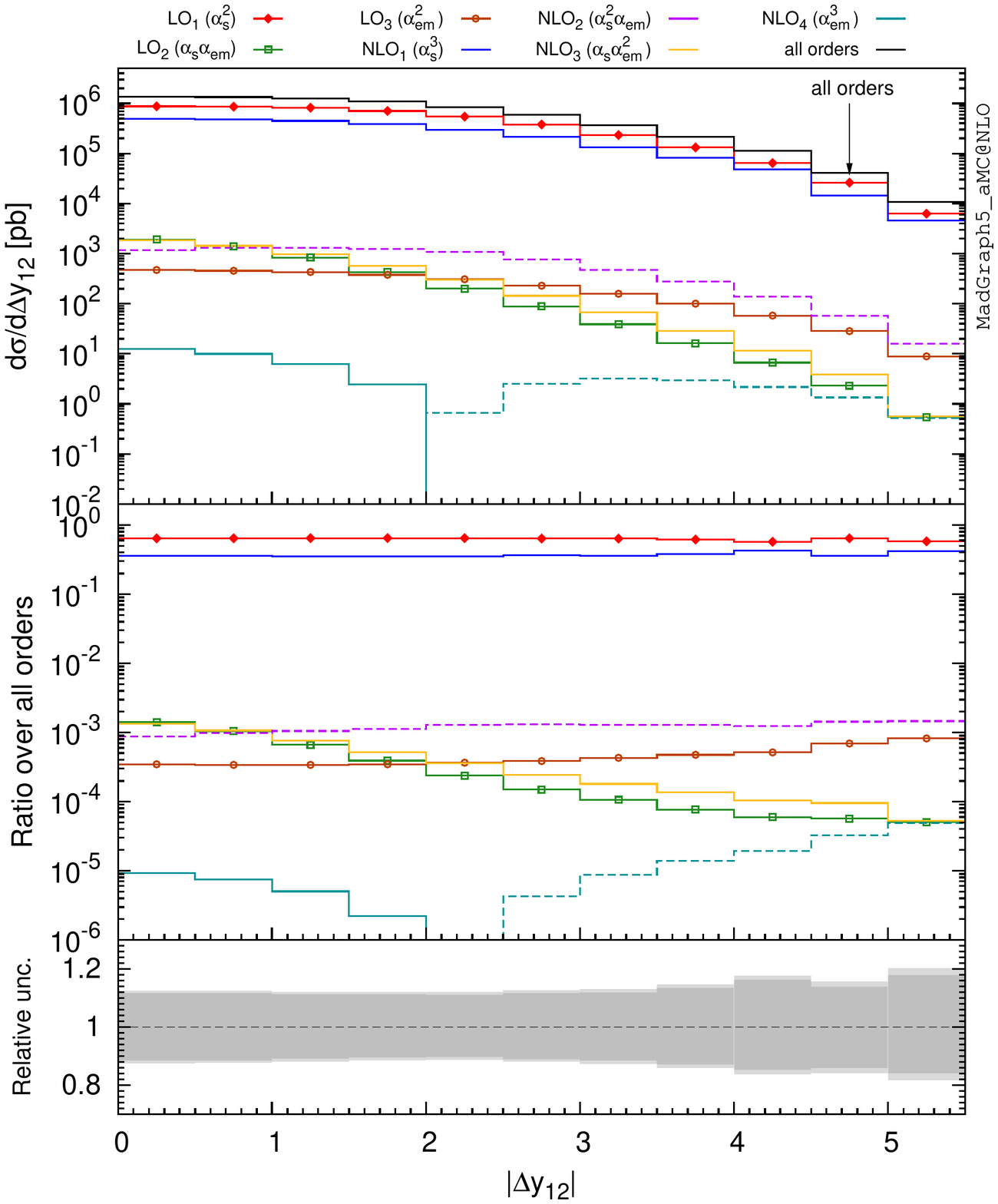}
\end{center}
\vskip -3.5truecm
\caption{\label{fig:Dy1} 
Rapidity distance between the two hardest jets.
}
\end{figure}
\begin{figure}[!ht]
\vskip -1.0truecm
\begin{center}
  \includegraphics[width=0.99\textwidth]{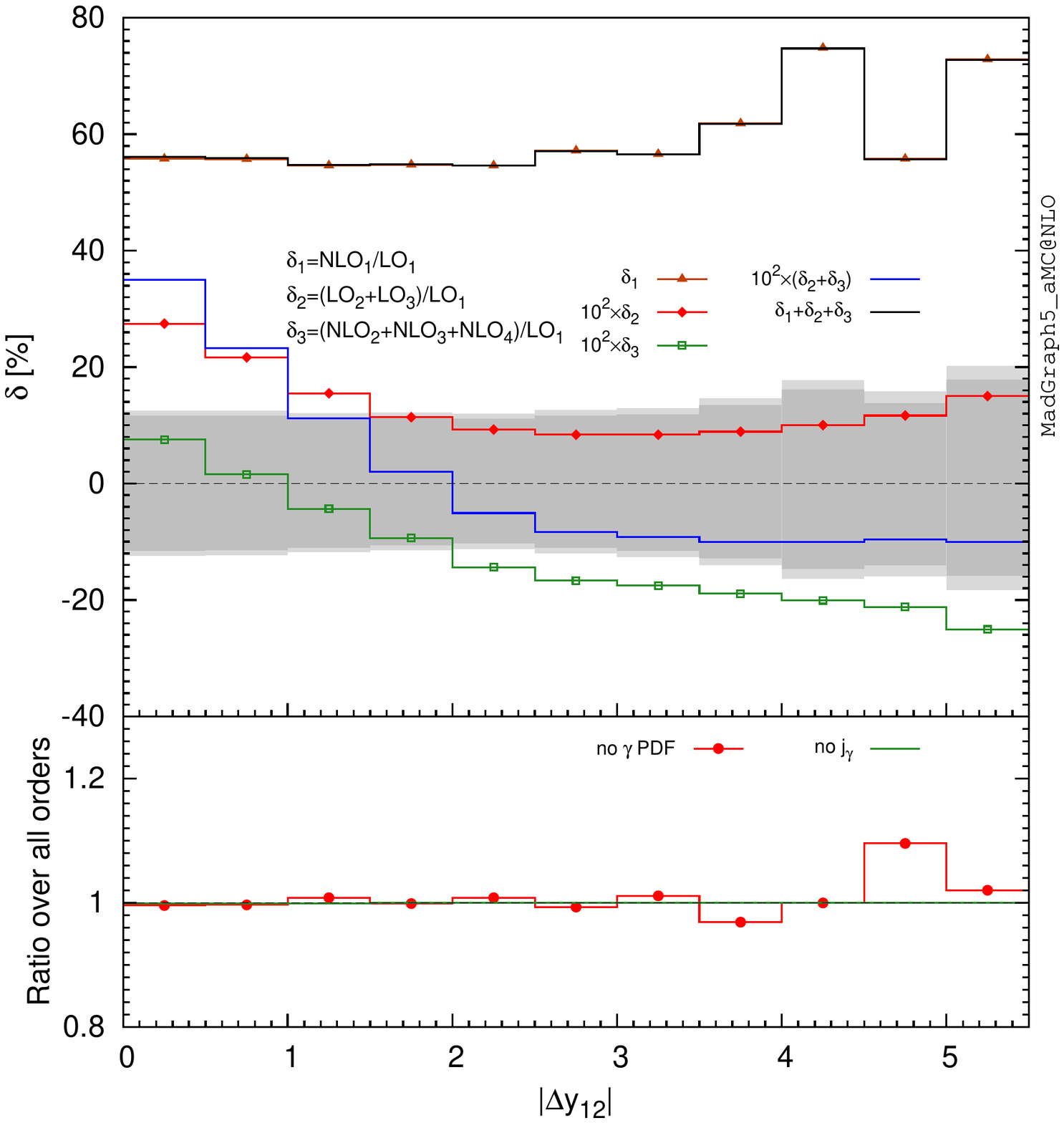}
\end{center}
\vskip -5.5truecm
\caption{\label{fig:Dy2} 
Rapidity distance between the two hardest jets.
}
\end{figure}

We now turn our attention to differential observables. We shall present
six of them in figs.~\ref{fig:pt1}--\ref{fig:Dy2}, with two figures for
each observable (plus one relevant to the direct comparison of $\pt$
results in different rapidity ranges, fig.~\ref{fig:pt2all}). 
The patterns in the layout of the plots are the same
for all of the observables; thus, we shall explain their meaning by using
the case of the single-inclusive jet transverse momentum $\ptinc$
(figs.~\ref{fig:pt1} and~\ref{fig:pt2}) in order to be definite.

There are three panels in fig.~\ref{fig:pt1}. The upper one presents the
absolute values of the three LO and the four NLO contributions to the cross
section, as well as their sum; as was previously mentioned, a solid (dashed)
pattern indicates that the corresponding result is positive (negative). The
three LO results are displayed as histograms overlaid with symbols: red with
full diamonds for $\Sigma_{\LOo}$, green with open boxes for $\Sigma_{\LOt}$,
and brown with open circles for $\Sigma_{\LOth}$. The four NLO results are
associated with plain histograms: blue for $\Sigma_{\NLOo}$, purple for
$\Sigma_{\NLOt}$, yellow for $\Sigma_{\NLOth}$, and cyan for $\Sigma_{\NLOf}$;
the sum of all contributions is represented by the black histogram. 
The middle inset presents the ratios of the
results shown in the upper inset, over the all-orders prediction; in other
words, these are the fractional contributions of the $\Sigma_{\LOi}$ and
$\Sigma_{\NLOi}$ terms to the most accurate result obtained from our
simulations.  The patterns employed in the middle inset are identical to those
of the upper inset. Finally, the bottom inset presents the relative
theoretical uncertainty of the all-orders result, in two different ways: the
light gray band corresponds to the hard-scale and PDF systematics (with the
two summed linearly), while the dark gray band shows the hard-scale
uncertainty only (see eq.~(\ref{scalevar}) and thereabouts).

The main frame of fig.~\ref{fig:pt2} presents various linear combinations 
of the results shown in fig.~\ref{fig:pt1}, in the form of ratios over
the leading LO prediction, $\Sigma_{\LOo}$. In particular, we have defined
the quantities:
\beqn
\delta_1&=&\frac{\Sigma_{\NLOo}}{\Sigma_{\LOo}}\,,
\\
\delta_2&=&\frac{\Sigma_{\LOt}+\Sigma_{\LOth}}{\Sigma_{\LOo}}\,,
\\
\delta_3&=&\frac{\Sigma_{\NLOt}+\Sigma_{\NLOth}+\Sigma_{\NLOf}}
{\Sigma_{\LOo}}\,,
\eeqn
which are displayed as a brown histogram overlaid with full triangles
($\delta_1$), a red histogram overlaid with full diamonds ($\delta_2$), and a
green histogram overlaid with open boxes ($\delta_3$), respectively.  We also
show the sum $\delta_2+\delta_3$ as a blue histogram, and the sum
$\delta_1+\delta_2+\delta_3$ as a black histogram. Finally, we report for
reference the two uncertainty bands already shown in fig.~\ref{fig:pt1}. In
view of the definition of $\Sigma_{\LOi}$ and $\Sigma_{\NLOi}$, the physical
meaning of the various curves presented in fig.~\ref{fig:pt2} is the
following. $\delta_1$ is equal to \mbox{$K_{\rm QCD}-1$}, with $K_{\rm QCD}$ 
the $K$ factor associated with a pure-QCD computation. $\delta_2$ measures 
the relative impact of the two Born contributions which are non pure-QCD.
$\delta_3$ is equal to \mbox{$K_{\cancel{\rm QCD}}-1$}, with 
$K_{\cancel{\rm QCD}}$ the $K$ factor associated with NLO contributions that 
are not pure QCD\footnote{As was already said, and for the sake of 
consistency, this $K$ factor is defined by using the pure-QCD Born 
$\Sigma_{\LOo}$ in the denominator.}.
Thus, $\delta_1+\delta_2+\delta_3$ shows the effect on the best (i.e.~the
all-orders one) prediction of all contributions different from the dominant
Born one ($\Sigma_{\LOo}$), while the comparison between $\delta_1$ and
$\delta_2+\delta_3$ allows an immediate understanding of how much of that is
due to either pure-QCD NLO corrections, or to other LO and NLO contributions.

The lower panel of fig.~\ref{fig:pt2} displays two results, both of which
are ratios of all-orders predictions obtained with specific conditions over
the all-orders default prediction. The red histogram overlaid with full 
circles corresponds to setting to zero the photon PDF, while the green
histogram corresponds to removing the photon-jet contributions.

The predictions for the single-inclusive jet transverse momentum 
shown in figs.~\ref{fig:pt1} and~\ref{fig:pt2} are dominated by the
leading contributions at both the LO and the NLO for $\ptinc\lesssim 2$~TeV.
The impact of non-QCD contributions is essentially negligible up to those
values, well within the scale uncertainty band. As is clear from 
fig.~\ref{fig:pt2}, specifically from the comparison of $\delta_2$,
$\delta_3$, and $\delta_2+\delta_3$, this is chiefly due to the very
large cancellation that occurs between the $\LOi$ and the $\NLOi$ 
terms ($i\ge 2$) -- note, from fig.~\ref{fig:pt1}, that this is not only true 
for the sums of such terms, but to some extent also for them individually,
since the NLO ones are negative either in all or in a large part of the
$\pt$ range considered. Eventually, the LO cross sections grow faster in
absolute value than their NLO counterparts. Thus, the sum of all results 
minus the leading LO term $\Sigma_{\LOo}$ is indistinguishable from 
$\Sigma_{\NLOo}$ up to 2~TeV, but then starts to differ significantly 
from it, to the extent that $\Sigma_{\NLOo}$ contributes to less than 50\% 
to the sum for those transverse momenta at the upper end of the range 
probed in our plots, $\ptinc\gtrsim 4.5$~TeV. When one moves towards
such large $\ptinc$'s, one sees that the NLO scale uncertainty remains
moderate, while that due to the PDFs grows rapidly, owing to the poor
constraining power of the data currently used in PDF fits on the
corresponding $x$ region. To that PDF uncertainty, the photon
contribution increases with $\ptinc$ (being equal to about 3\% of 
the total PDF uncertainty at $\ptinc\simeq 2.6$~TeV, and to about 22\% 
at $\ptinc\simeq 4.6$~TeV), but is never the dominant effect.
From fig.~\ref{fig:pt2} we see that the impact of the contributions
that depend on the photon PDF is negligible for $\ptinc\lesssim 3.5$~TeV,
while it becomes substantial for larger values of the transverse momentum. 
Needless to say, the validity of this observation is restricted to the PDF 
used in the present simulations. The photon component in the NNPDF2.3QED set 
is mainly constrained by LHC Drell-Yan data via a reweighting procedure. This 
results in a significant photon density at large $x$ that, however, is 
associated with 
a sizeable uncertainty. Other approaches, which rely either on assumptions on 
the functional form at some initial scale~\cite{ Martin:2004dh,Schmidt:2015zda,
Harland-Lang:2016kog}, or on a direct extraction from proton structure 
functions~\cite{Manohar:2016nzj}, suggest that its central value is much 
smaller than the NNPDF2.3 one at large $x$ and rather precisely determined
(in the recent sets), thus effectively lying close to the lower limit of 
the NNPDF2.3QED uncertainty band.

We also remark that the removal of the photon-jet cross sections has
a negligible impact in the whole transverse momentum range considered.
It does affect the individual $\LOi$ and $\NLOi$, $i\ge 2$ contributions,
especially $\LOt$ where it can be as large as  30\%; however, this occurs
mostly for $\ptinc\lesssim 0.5$~TeV, where non-QCD terms can be safely ignored.

The single-inclusive transverse momentum is again shown in
figs.~\ref{fig:pt1A} and~\ref{fig:pt2A}, subject to the constraint
$\abs{y}\le 1$ (in other words, each jet that gives a contribution
to these histograms must satisfy a small-rapidity constraint). The 
patterns in these figures are very similar to those of figs.~\ref{fig:pt1} 
and~\ref{fig:pt2}, respectively, owing to the dominance of central
jets in the case inclusive over the whole rapidity range.

Things slightly change when one considers the rapidity intervals 
$1<\abs{y}\le 2$ and $2<\abs{y}\le 2.8$, whose cases are presented in 
figs.~\ref{fig:pt1B} and~\ref{fig:pt2B}, and in figs.~\ref{fig:pt1C} 
and~\ref{fig:pt2C}, respectively. In the transverse momentum region
$\ptinc\lesssim 1$~TeV, $\delta_2$ tends to be marginally flatter
when the rapidity is increased; conversely, $\delta_3$ decreases, somehow
more rapidly. The net effect is that the amount of cancellation between 
the LO and NLO cross sections is smaller the farther away one moves from 
central rapidities in this range of relatively small $\ptinc$'s, so that
the overall EW effects, that decrease the pure-QCD cross sections, are
stronger the larger the rapidities. This is seen more clearly in 
fig.~\ref{fig:pt2all}, where the results for $\delta_2$, $\delta_3$, 
and $\delta_2+\delta_3$, already shown in figs.~\ref{fig:pt2A}, \ref{fig:pt2B}, 
and~\ref{fig:pt2C}, are presented together (as red, green, and blue histograms;
the $\abs{y}\le 1$, $1<\abs{y}\le 2$, and $2<\abs{y}\le 2.8$ predictions
are displayed as solid, dashed, and short-dashed histograms, respectively),
by using a smaller $y$-axis scale w.r.t.~those of the original plots. 
For larger transverse momenta the trend changes, with the positive
LO contributions eventually becoming larger than their NLO counterparts
(in absolute value). Thus, the $\delta_2+\delta_3$ prediction crosses zero 
at $\ptinc\sim 1.6$~TeV for $\abs{y}\le 1$, and at $\ptinc\sim 2.5$~TeV for 
$1<\abs{y}\le 2$ (the statistics is insufficient to draw any conclusion 
in the range $2<\abs{y}\le 2.8$).

We conclude that, as far as the single-inclusive transverse momentum
is concerned, the impact of LO and NLO contributions beyond the leading 
ones do depend on the rapidity range considered, and tends to decrease
(increase) the pure-QCD results when moving away from the central region 
for small (large) $\ptinc$; in all cases, the absolute values of the
overall effects are relatively small. This pattern is due to a variety
of reasons; in particular, one may mention the fact that, the larger the 
rapidity, the more difficult it is to reach the high-$\pt$ region where 
EW effects are known to be more prominent, but also the fact that the 
extent of the cancellation between LO and NLO results is difficult to
be predicted {\em a priori}. In any case, such a pattern must be taken 
into account in the context of PDF fits that aim to include EW corrections, 
and that need to consider different rapidity ranges in order to constrain 
more effectively the small-$x$ region.

Our predictions for the invariant mass of the hardest-jet pair are
given in figs.~\ref{fig:M121} and~\ref{fig:M122} (note that some of
the histograms have been rescaled in the latter figure, in order to 
make them more clearly visible in the layout). NLO corrections
are dominated by the pure-QCD ones $\Sigma_{\NLOo}$, that turn negative
around $M_{12}\simeq 1$~TeV\footnote{There is a visible numerical
instability that affects the large-mass predictions of $\Sigma_{\NLOo}$. It 
is due to significant cancellations between the real-emission and virtual 
contributions to that mass region.}. EW effects tend to decrease the cross 
section further, with the second-leading NLO corrections $\Sigma_{\NLOt}$ 
being negative and larger in absolute value than the second-leading LO term
$\Sigma_{\LOt}$. However, the overall impact on the physical cross section
is rather small, and in particular smaller than the hard-scale uncertainty.
As was observed in ref.~\cite{Dittmaier:2012kx}, even for mass values 
of several TeV's one is not fully in the Sudakov region, and thus EW 
contributions tend to follow the hierarchy established by the couplings,
without major logarithmic enhancements. We also observe a very small impact
of the removal of the photon jets. In this regard, the same comments as for 
the single-inclusive transverse momentum apply here. By removing photon-jet
cross sections from $\Sigma_{\LOt}$, that term is halved at invariant
masses smaller than 0.5~TeV; however, as can be seen from fig.~\ref{fig:M121},
in that region its contribution to the all-orders rate is in practice 
negligible.

We finally show, in figs.~\ref{fig:Dy1} and~\ref{fig:Dy2}, the rapidity
separation between the two hardest jets (again, some of the histograms 
have been rescaled in the latter plot to improve its readability). This
observable is dominated by low-$\pt$ configurations, and as a consequence
of that subleading terms, both at the LO and the NLO, are numerically
extremely small, and completely swamped by hard-scale uncertainties.
Leading NLO corrections are large, but almost flat in the whole
range considered. As in the previous cases, the removal of photon jets
is irrelevant to the all-orders result, while being important up to the
largest rapidity separations in particular for $\Sigma_{\LOt}$.

\section{Conclusions\label{sec:conc}}
In this paper we have studied the hadroproduction of dijets, and
considered all of the LO and NLO contributions of QCD and EW origin
to the corresponding cross section, presented as single-inclusive
distributions and two-jet correlations for $pp$ collisions at 13~TeV.
By doing so, we have computed for the first time three subleading
NLO corrections: the ${\cal O}(\as^2\aem)$ electromagnetic one
(our results include the contributions due to real-photon emissions),
and the ${\cal O}(\as\aem^2)$ and ${\cal O}(\aem^3)$ EW ones.
The calculations have been performed in the automated \aNLO\ framework,
which is thus extensively tested in a mixed-coupling scenario that 
features both EW and QCD loop corrections, and both QCD and QED
real-emission subtractions.

When all subleading NLO corrections are computed, it is necessary
to be particularly careful in the case one wants to not take into
account jets that are predominantly of electromagnetic origin.
Although from the phenomenological viewpoint we do not consider
this operation to have a compelling motivation, we have outlined 
an IR-safe scheme through which this result can be achieved. 
Its exact implementation requires the use of fragmentation functions,
whose determination from data is either poor or not available at 
present\footnote{However, the necessary ingredients for a technically-viable
computation that leads to IR-finite cross sections can all be derived
from purely perturbative information.}.
For the sake of this paper, we have adopted a more pragmatic strategy, 
which is a (perturbative) approximation of the more general scheme, that
does not employ the fragmentation functions. We have shown that the
removal of EW-dominated jets has a negligible impact at the level
of observable differential rates, and one can thus safely work 
with democratic jets, in which all massless particles (quarks,
gluons, photons, and leptons) are treated on equal footing.

In general, contributions that are expected to be subleading
according to the coupling-constant combination they feature turn
out to be indeed numerically subleading, with pure-QCD effects being
dominant everywhere, except in the very-high transverse momentum
region of the single-inclusive jet $\pt$. In other words, within the LO 
and NLO cross sections, we find that the hierarchy naively established 
on the basis of the couplings is largely respected, but we also 
remark that, in a significant fraction of the phase space,
$\Sigma_{\NLOt}$ is larger than $\Sigma_{\LOt}$. For all observables 
considered here, there are large cancellations between the LO and
NLO subleading terms, which is one of the major motivations for
computing them all in a consistent manner.

\section*{Acknowledgements}
This work is supported in part by the ERC grant 291377
``LHCtheory: Theoretical predictions and analyses of LHC physics: advancing
the precision frontier''. SF thanks the CERN TH division for hospitality
during the course of this work. SF is indebted to Stefano Catani for
many stimulating discussions, and for comments on the manuscript; he is 
also grateful to Fabio Cossutti, Hannes Jung, Andrew Larkoski, and 
Wouter Waalewijn for comments on different aspects of their work.
The work of MZ is supported by the
European Union's Horizon 2020 research and innovation
programme under the Marie Sklodovska-Curie grant agreement No 660171 and in
part by the ILP LABEX (ANR-10-LABX-63), in turn supported by French state funds
managed by the ANR within the ``Investissements d'Avenir'' programme under
reference ANR-11-IDEX-0004-02.
RF and DP are supported by the Alexander von Humboldt Foundation in the
framework of the Sofja Kovalevskaja Award Project ``Event Simulation for the
Large Hadron Collider at High Precision''.
The work of VH is supported by the Swiss National Science Foundation (SNSF)
with grant PBELP2 146525.

\phantomsection
\addcontentsline{toc}{section}{References}
\bibliographystyle{JHEP}
\bibliography{jjrefs}

\end{document}